# Constraints, observations, and experiments on the simulation hypothesis

Florian Neukart
Terra Quantum AG
University of Leiden
f.neukart@terraquantum.swiss

Anders Indset
Philosopher
ai@andersindset.com

Markus Pflitsch
Terra Quantum AG
pflitsch@terraquantum.swiss

Michael R. Perelshtein
Terra Quantum AG
Aalto University
mpe@terraquantum.swiss

*Abstract*—The question "What is real?" can be traced back to the shadows in Plato's cave. Two thousand years later, René Descartes lacked knowledge about arguing against an evil deceiver feeding us the illusion of sensation. Descartes' epistemological concept later led to various theories of what our sensory experiences actually are. The concept of "illusionism", proposing that even the very conscious experience we have – our qualia – is an illusion, is not only a red-pill scenario found in the 1999 science fiction movie "The Matrix" but is also a philosophical concept promoted by modern tinkers, most prominently by Daniel Dennett. He describes his argument against qualia as materialistic and scientific. Reflection upon a possible simulation and our perceived reality was beautifully visualized in "The Matrix", bringing the old ideas of Descartes to coffee houses around the world. Irish philosopher Bishop Berkeley was the father of what has later been coined as "subjective idealism", basically stating that "what you perceive is real" (e.g., "The Matrix" is real because its population perceives it). Berkeley then argued against Isaac Newton's absolutism of space, time, and motion in 1721, ultimately leading to Ernst Mach and Albert Einstein's respective views. Several neuroscientists have rejected Dennett's perspective on the illusion of consciousness, and idealism is often dismissed as the notion that people want to pick and choose the tenets of reality. Even Einstein ended his life on a philosophical note, pondering the very foundations of reality. With the advent of quantum technologies based on the control of individual fundamental particles, the question of whether our universe is a simulation isn't just intriguing. Our ever-advancing understanding of fundamental physical processes will likely lead us to build quantum computers utilizing quantum effects for simulating nature quantum-mechanically in all complexity, as famously envisioned by Richard Feynman. Finding an answer to the simulation question will potentially alter our very definition and understanding of life, reshape theories on the evolution and fate of the universe, and impact theology. No direct observations provide evidence in favor or against the simulation hypothesis, and experiments are needed to verify or refute it. In this paper, we outline several constraints on the limits of computability and predictability in/of the universe, which we then use to design experiments allowing for first conclusions as to whether we participate in a simulation chain. We elaborate on how the currently understood laws of physics in both complete and small-scale universe simulations prevent us from making predictions relating to the future states of a universe, as well as how every physically accurate simulation will increase in complexity and exhaust computational resources as global thermodynamic entropy increases. Eventually, in a simulation in which the computer simulating a universe is governed by the same physical laws as the simulation and is smaller than the universe it simulates, the exhaustion of computational resources will halt all simulations down the simulation chain unless an external programmer intervenes or isn't limited by the simulation's physical laws, which we may be able to observe. By creating a simulation chain and observing the evolution of simulation behavior throughout the hierarchy taking stock of statistical relevance, as well as comparing various least complex simulations under computability and predictability constraints, we can gain insight into whether our universe is part of a simulation chain.

## I. Introduction

A dream within a dream from the ancient philosophical thinking of inception is nothing but today's reflection of the simulation within a simulation. Over the past years, the question of reality has gained massive mainstream media attention as prominent figures like Elon Musk have stated there is a "one in billions" chance we do not live in a simulation [1], and pop star astrophysicist Neil deGrasse Tyson has also jumped onto the idea stating that the probability is more than 50% [2]. In addition, philosopher David Chalmers has also caught on to the belief that we likely live in a simulation [3,4], pushing for further examination of the very notion.

However realistic or plausible such a hypothesis [5, 6] may be, how could modern physics and mathematics support seeking evidence for such a case? Scientists have criticized the hypothesis made by philosopher Bostrom for being pseudoscience [7, 8] as it sidesteps the current laws of physics and lacks a fundamental understanding of general relativity. Suppose an external programmer - an entity running a simulation and characterized as external to the simulation - could define the simulation's physical laws. What would an external programmer and beings within the simulation be able to calculate based on their understanding of physical laws? Moreover, theoretically or practically, could beings in the simulation conceive and implement the apparatus or tools to verify that they aren't participating in a simulation chain?

While controversial, the question of whether we exist in a simulation and thus participate in a simulation chain cannot be answered with certainty today. Nevertheless, it is intriguing, as an answer to it could lead us to question our very definitions of life and spirituality. Suppose we spark a chain of simulations, each hosting intelligent life intending to simulate the universe. Would we classify each of the simulated life forms as actual life? What if we could confidently state that we are part of a simulation chain and simulated beings ourselves? Would that change our definition of what counts as "real" or "artificial" life? In the argument made by Bostrom, one premise is worth examination: if there is a physical possibility of creating a simulation, then based on the state of development and the relation to time access, there would most likely be a higher probability of our residing within such a simulation than our being the exact generation building such a simulation. Experiments are needed to gain deeper insights, but several constraints prevent us from designing experiments that directly answer the question of whether an external programmer has created the universe and whether it's only one of infinite hierarchical simulation chains. However, it is possible to indirectly test the simulation hypothesis under certain assumptions. The outlined experiments for doing so involve creating a simulation, potentially resulting in a chain of simulations, and conducting observations on the simulation behavior within the confines of a hierarchy until statistical relevance can be obtained. Potential observations of note could include the emergence of intelligent life and its behavior, a reversal of global entropy, compactification of dimensions, or the evolution of simulations along the simulation chain (all of which are, based on the current understanding of physics, impossible for us to conduct in our universe, but an external programmer shall not suffer from such limitations). Designing such experiments leads to the ultimate boundaries of computability and predictability. Physical and computational constraints prevent us from simulating a universe equal in complexity and size to our universe and from making accurate predictions of the future, whether or not the "real" or simulated universes are based on the same physical laws.

Moreover, the cosmos has yet to be fully understood. For example, the universe's fate and how to unite quantum physics and general relativity are deep and open questions. Today, quantum theory is widely understood as an incomplete theory, and new models may be discovered that will further flesh out our understanding of what quantum theory has indicated thus far. However, the state of modern physics and our imagination allows us to conceive experiments and build advanced technologies to continue scientific progress; therefore, the current framework shall not hold us back from searching for evidence related to the simulation hypothesis. The entrance point, however, must be the current understanding of mathematics and the challenges associated with our current knowledge of physics. Therefore, conducting experiments on such a hypothesis naturally requires assumptions to be made.

Also, many open questions remain in living systems theory, and we don't yet know with certainty whether or not we are the only intelligent species in the universe. Still, we can conceive experiments that help us to gain insights into the ultimate questions: Were our universe and everything in it created, or did it emerge by itself? Is our universe unique, or is it just one of many, as described by the many-worlds interpretation of quantum physics [9]? In the article, we outline some fundamentals of computing and physics, which will help us define the experiment's constraints. First, quantum physics is the essential pillar we build our experiments on - ergo, the current understanding of quantum mechanics - as our current understanding constitutes the most fundamental physics in the universe that everything else is based upon. Secondly, we briefly introduce different fates of the universe that the scientific community assumes to be scientifically sound and further guide us in designing an experiment independent of how the universe evolves. Thirdly, we consider the ultimate limits of computability, which also lead us back to quantum physics, both when it comes to engineering quantum computers and simulating physical and chemical processes in the universe. While Alan Turing showed what is computable [10], we indicate which computers are constructible within this universe. Finally, we explore different interpretations of observations gained from simulation chains and individual specimens we base the proposed experiments on. We also investigate observations in our universe, indicating whether we participate in a simulation chain.

## II. THE SIMULATION HYPOTHESIS

The simulation hypothesis, first proposed by philosopher Nick Bostrom in 2003 [5, 6], is the consequence of an assumption in a thought model, which is sometimes also called "the simulation argument" [3,5,6]. It consists of three alternatives to the real or simulated existence of developed civilizations, at least one of which is said to be true. According to the simulation hypothesis, most contemporary humans are simulations, not actual humans. The simulation hypothesis is distinguished from the simulation argument by allowing this single assumption. It is no more likely or less likely than the other two possibilities of the simulation argument. In a conceptual model in the form of an OR link, the following three basic possibilities of technically "immature" civilizations – like ours - are assumed. At least one of the above possibilities should be true. A mature or post-human civilization is defined as one that has the computing power and knowledge to simulate conscious, self-replicating beings at a high level of detail (possibly down to the molecular nanobot level). Immature civilizations do not have this ability. The three choices are [5]:

1) Human civilization will likely die out before reaching a post-human stage. If this is true, then it almost certainly follows that human civilizations at our level of technological development will not reach a post-human level.
2) The proportion of post-human civilizations interested in running simulations of their evolutionary histories, or variations thereof, is close to zero. If this is true, there is a high degree of convergence among technologically advanced civilizations. None of them contain individuals interested in running simulations of their ancestors (ancestor simulations).
3) We most likely live in a computer simulation. If this is true, we almost certainly live in a simulation, and most people do. All three possibilities are similarly likely. If we don't live in a simulation today, our descendants are less likely to run predecessor simulations. In other words, the belief that we may someday reach a posthuman level at which we run computer simulations is wrong unless we already live in a simulation today.

According to the simulation hypothesis, at least one of the three possibilities above is true. It is argued on the additional assumption that the first two possibilities do not occur, for example, that a considerable part of our civilization achieves technological maturity and, secondly, that a significant amount of civilization remains interested in using the resources to develop predecessor simulations. If this is true, the size of the previous simulations reaches astronomical numbers in a technologically mature civilization. This happens based on an extrapolation of the high computing power and its exponential growth, the possibility that billions of people with their computers can run previous simulations with countless simulated agents, as well as from technological progress with some adaptive artificial intelligence, what an advanced civilization possesses and uses, at least in part, for predecessor simulations. The consequence of the simulation of our existence follows from the assumption that the first two possibilities are incorrect. There are many more simulated people like us in this case than non-simulated ones. For every historical person, there are millions of simulated people. In other words, almost everyone at our experience level is likelier to live in simulations than outside of them [3]. The conclusion of the simulation hypothesis is described from the three basic possibilities and from the assumption that the first two possibilities are not true as the structure of the simulation argument. The simulation hypothesis that humans are simulations does not follow the simulation argument. Instead, the simulation argument shows all three possibilities mentioned side by side, one of which is true. But it remains to be seen what that is. It is also possible that the first assumption will come true, according to which all civilizations and, thus, humankind will die out for some reason. According to Bostrom, there is no evidence for or against accepting the simulation hypothesis that we are simulated beings, nor the correctness of the other two assumptions [5].

From a scientific standpoint, everything in our perceived reality could be coded out as the foundation of the scientific assumption that the laws of nature are governed by mathematical principles describing some physicality. The fact that an external programmer can control the laws of physics and even play with them has been deemed controversial in the simulation hypothesis. Something "outside of the simulation" an external programmer - is, therefore, more of a sophisticated and modern view of the foundation of monotheistic religions/belief systems. Swedish techno-philosopher Alexander Bard proposed that the theory of creationism be moved to physics [11], and the development of super (digital) intelligence was the creation of god, turning the intentions of monotheism from the creator to the created. Moving from faith and philosophical contemplation towards progress in scientific explanation is what the advancement of quantum technology might propose.

The critics of Bostrom state that we do not know how to simulate human consciousness [12–14]. An interesting philosophical problem here is the testability of whether a simulated conscious being – or uploaded consciousness – would remain conscious. The reflection on a simulated super intelligence without perception of its perception was proposed as a thought experiment in the "final narcissistic injury" (reference). Arguments against that include that with complexity, consciousness arises – it is an emergent phenomenon. A counterargument could easily be given that there seem to be numerous complex organs that seem unconscious, and also – despite reasoned statements by a former Google engineer [15] – that large amounts of information give birth to consciousness. With the rising awareness of the field, studies on quantum physical effects in the brain have also gained strong interest. Although rejected by many scientists, prominent thinkers such as Roger Penrose and Stuart Hameroff have proposed ideas around quantum properties in the brain [16]. Even though the argument has gained some recent experimental support [17], it is not directly relevant to the proposed experiments. A solution to a simulated consciousness still seems far away, even though it belongs to the seemingly easy problems of consciousness [18]. The hard problem of consciousness is why humans perceive to have phenomenal experiences at all [18]. Both don't tackle the meta-problem of consciousness stating why we believe that is a problem, that we have an issue with the hard problem of consciousness.

German physicist Sabine Hossenfelder has argued against the simulation hypothesis, stating it assumes we can reproduce all observations not employing the physical laws that have been confirmed to high precision but a different underlying algorithm, which the external programmer is running [19]. Hossenfelder does not believe this was what Bostrom intended to do, but it is what he did. He implicitly claimed that it is easy to reproduce the foundations of physics with something else. We can approximate the laws we know with a machine, but if that is what nature worked, we could see the difference. Indeed, physicists have looked for signs that natural laws proceed step-by-step, like a computer code. But their search has come up empty-handed. It is possible to tell the difference because attempts to algorithmically reproduce natural laws are usually incompatible with the symmetries with Einstein's Theories of Special and general relativity. Hossenfelder has stated that it doesn't help if you say the simulation would run on a quantum computer: "Quantum computers are special purpose machines. Nobody really knows how to put general relativity on a quantum computer" [19]. Hossenfelders criticism of Bostrom's argument continues with the statement that for it to work, a civilization needs to be able to simulate a lot of conscious beings. And, assuming they would be conscious beings, they would again need to simulate many conscious beings. That means the information we think the universe contains would need to be compressed. Therefore, Bostrom has to assume that it is possible to ignore many of the details in parts of the universe no one is currently looking at and then fill them in case someone looks. So, again, there is a need to explain how this is supposed to work. Hossenfelder asks the question what kind of computer code can do that? What algorithm can identify conscious subsystems and their intentions and quickly fill in the information without producing an observable inconsistency?

According to Hossenfelder, this is a much more critical problem than it seems Bostrom appreciates. She further states that one can't generally ignore physical processes on a short distance and still get the large distances right. Climate models are examples of this - with the currently available computing power models with radii in the range of tens of kilometers can be computed [20]. We can't ignore the physics below this scale, as the weather is a nonlinear system whose information from the short scales propagates to large scales. If short-distance physics can't be computed, it has to be replaced with something else. Getting this right, even approximately, is difficult. The only reason climate scientists get this about right is that they have observations that they can use to check whether their approximations work. Assuming the external programmer only has one simulation, like in the simulation hypothesis, there is a catch, as the external programmer would have to make many assumptions about the reproducibility of physical laws using computing devices. Usually, proponents don't explain how this is supposed to work. But finding alternative explanations that match all our observations to high precision is difficult. The simulation hypothesis, in its original form, therefore, isn't a serious scientific argument. That doesn't mean it is necessarily incorrect, but it requires a more solid experimental and logical basis instead of faith.

### III. QUANTUM PHYSICS

As Richard Feynman famously said, if we intend to simulate nature, we have to do it quantum mechanically, as nature is not classical.[1] While the transition dynamics from the microscopic to the macroscopic is not yet fully understood in every aspect, theory and experiments agree that macroscopic behavior can be derived from interactions at the quantum scale. Quantum physics underlies the workings of all fundamental particles; thus, it governs all physics and biology on larger scales. The quantum field theories of three out of four forces of nature, the weak nuclear force and the electromagnetic force [22,23], and the strong nuclear force [24] have been confirmed experimentally numerous times and have strongly contributed to the notion that quantum physics comprises, as of our current understanding, the most fundamental laws of nature. Immense efforts worldwide are underway to describe gravity quantum-mechanically [25–27], which has proven elusive. Gravity differs from the other interactions because it is caused by objects curving space-time around them instead of particle exchange. Uniting quantum physics with general relativity has proven to be one of the most formidable challenges in physics and our understanding of the universe [28,29]. Despite many scientific hurdles still to take, we have gained some insights into how the universe works. When we look at quantum physics as the "machine language of the universe", the universal interactions can be interpreted as higher-level programming languages. Quantum physics includes all phenomena and effects based on the observation that certain variables cannot assume any value but only fixed, discrete values. That also includes the wave-particle duality, the non-determination of physical processes, and their unavoidable influence by observation. Quantum physics includes all observations, theories, models, and concepts that go back to Max Planck's quantum hypothesis, which became necessary around 1900 because classical physics reached its limits, for example, when describing light or the structure of matter. The differences between quantum physics and classical physics are particularly evident on the microscopic scale, for example, the structure of atoms and molecules, or in particularly pure systems, such as superconductivity and laser radiation. Even the chemical or physical properties of different substances, such as color, ferromagnetism, electrical conductivity, etc., can only be understood in terms of quantum physics. Theoretical quantum physics includes quantum mechanics, describing the behavior of quantum objects under the influence of fields, and quantum field theory, which treats the fields as quantum objects. The predictions of both theories agree extremely well with the experimental results, and macroscopic behavior can be derived from the smallest scale. If we define reality as what we can perceive, detect, and measure around us, then quantum physics is the fabric of reality. Therefore, an accurate simulation of the universe, or parts of it, must have quantum physics as a foundation. The internal states of a computer used for simulation must be able to accurately represent all external states, requiring a computer that uses quantum effects for computation and can accurately mimic the behavior of all quantum objects, including their interactions. The requirements for such a computer go beyond the quantum computers built today and envisioned for the future. Engineering such a computer is a formidable challenge, which will be discussed in the following chapters.

One of the arguments presented later in this article is the physical predictability constraint, which prevents us from building a computer that can be used to predict any future states of the universe through simulation. Would nature be purely classical a computer would not suffer from that constraint (there are others, though), but quantum physics imposes some restrictions, no matter how advanced our theories on how nature works become. Within the framework of classical mechanics, the trajectory of a particle can be calculated entirely from its location and velocity if the acting forces are known. The state of the particle can thus be described unequivocally by two quantities, which, in ideal measurements, can be measured with unequivocal results. Therefore, a separate treatment of the state and the measured variables or observables is not necessary for classical mechanics because the state determines the measured values and vice versa. However, nature shows quantum phenomena that these terms cannot describe. On the quantum scale, it is no longer possible to predict where and at what speed a particle will be detected. If, for example, a scattering experiment with a particle is repeated under precisely the same initial conditions, the same state must always be assumed for the particle after the scattering process, although it can hit different places on the screen. The state of the particle after the scattering process does not determine its flight direction. In general, there are states in quantum mechanics that do not allow the prediction of a single measurement result, even if the state is known exactly. Only probabilities can be assigned to the potentially measured values. Therefore, quantum mechanics treats quantities and states separately, and different concepts are used for these quantities than in classical mechanics.

---

[1] "Nature isn't classical, dammit, and if you want to make a simulation of nature, you'd better make it quantum mechanical, and by golly, it's a wonderful problem because it doesn't look so easy." [21]

In quantum mechanics, all measurable properties of a physical system are assigned mathematical objects, the so-called observables. Examples are the location of a particle, its momentum, its angular momentum, or its energy. For every observable, there is a set of special states in which the result of a measurement cannot scatter but is clearly fixed. Such a state is called the eigenstate of the observable, and the associated measurement result is one of the eigenvalues of the observable. Different measurement results are possible in all other states that are not an eigenstate of this observable. What is certain, however, is that one of the eigenvalues is determined during this measurement and that the system is then in the corresponding eigenstate of this observable. For determining which of the eigenvalues is to be expected for the second observable or - equivalently - in which state the system will be after this measurement, only a probability distribution can be given, which can be determined from the initial state. In general, different observables have different eigenstates. For a system assuming the eigenstate of one observable as its initial state, the measurement result of a second observable is indeterminate. The initial state is interpreted as a superposition of all possible eigenstates of the second observable. The proportion of a certain eigenstate is called its probability amplitude. The square of the absolute value of a probability amplitude indicates the probability of obtaining the corresponding eigenvalue of the second observable in a measurement at the initial state. In general, any quantum mechanical state can be represented as a superposition of different eigenstates of an observable. Different states only differ in which of these eigenstates contribute to the superposition and to what extent.

Only discrete eigenvalues are allowed for some observables, such as angular momentum. In the case of the particle location, on the other hand, the eigenvalues form a continuum. The probability amplitude for finding the particle at a specific location is therefore given in the form of a location-dependent function, the so-called wave function. The square of the absolute value of the wave function at a specific location indicates the spatial density of the probability of finding the particle there.

Not all quantum mechanical observables have a classical counterpart. An example is spin, which cannot be traced back to properties known from classical physics, such as charge, mass, location, or momentum. In quantum mechanics, the description of the temporal development of an isolated system is analogous to classical mechanics employing an equation of motion, the Schrodinger equation. By solving this differential equation, one can calculate how the system's wave function evolves (see Eq. 1).

$$i\hbar \frac{\partial}{\partial t}\psi = \widehat{H}\psi \qquad (1)$$

In Eq. 1, the Hamilton operator $\widehat{H}$ describes the energy of the quantum mechanical system. The Hamilton operator consists of a term for the kinetic energy of the particles in the system and a second term that describes the interactions between them in the case of several particles and the potential energy in the case of external fields, whereby the external fields can also be time-dependent. In contrast to Newtonian mechanics, interactions between different particles are not described as forces but as energy terms, similar to the methodology of classical Hamiltonian mechanics. Here, the electromagnetic interaction is particularly relevant in the typical applications to atoms, molecules, and solids.

The Schrödinger equation is a first-order partial differential equation in the time coordinate, so the time evolution of the quantum mechanical state of a closed system is entirely deterministic. If the Hamilton operator $\widehat{H}$ of a system doesn't itself depend on time, this system has stationary states, i.e., states that do not change over time. They are the eigenstates of the Hamilton operator $\widehat{H}$. Only in them does the system have a well-defined energy $E$, for example, the respective eigenvalue (see Eq. 2).

$$\widehat{H}\psi = E\psi \qquad (2)$$

The Schrödinger equation then reduces to Eq. 3

$$i\hbar \frac{\partial}{\partial t}\psi = E\psi \qquad (3)$$

Quantum mechanics also describes how accurately we can measure and, thus, how accurately we can make predictions. Niels Bohr famously complained that predictions are hard, especially about the future. The uncertainty principle of quantum mechanics, which is known in the form of Heisenberg's uncertainty principle, relates the smallest possible theoretically achievable uncertainty ranges of two measurands. It is valid for every pair of complementary observables, particularly for pairs of observables which, like position and momentum or angle of rotation and angular momentum, describe physical measurands, which in classical mechanics are called canonically conjugate and which can assume continuous values.

If one of these quantities has an exactly determined value for the system under consideration, then the value of the others is entirely undetermined. However, this extreme case is only of theoretical interest because no real measurement can be entirely exact. In fact, the final state of the measurement of the observable $A$ is therefore not a pure eigenstate of the observable $A$, but a superposition of several of these states to a certain range of eigenvalues to $A$. If $\Delta A$ is used to denote the uncertainty range of $A$, mathematically defined by the so-called standard deviation, then the uncertainty range $\Delta B$ of the canonical conjugate observable $B$ the inequality in Eq. 4 is valid.

$$\Delta A \cdot \Delta B \geq \frac{h}{4\pi} = \frac{\hbar}{2} \qquad (4)$$

Another quantum-physical phenomenon is entanglement: a composite physical system, for example, a system with several particles, viewed as a whole, assumes a well-defined state without being able to assign a well-defined state to each subsystem. This phenomenon cannot exist in classical physics. There, composite systems are always separable. That is, each subsystem has a specific state at all times that determines its individual behavior, with the totality of the states of the individual subsystems and their interaction fully explaining the behavior of the overall system. In a quantum-physically entangled state of the system, on the other hand, the subsystems have several of their possible states next to each other, with each of these states of a subsystem being assigned a different state of the other subsystems. To explain the overall system's behavior correctly, one must consider all these coexisting possibilities together. Nevertheless, when a measurement is carried out on each subsystem, it always shows only one of these possibilities, with the probability that this particular result occurs being determined by a probability distribution. Measurement results from several entangled subsystems are correlated with one another; that is, depending on the measurement result from one subsystem, there is a different probability distribution for the possible measurement results from the other subsystems.

There is a lot more to say about quantum physics and how it is different from the everyday macroscopic world that we perceive, but suffice to say, both entanglement and superposition already massively add to the complexities of a simulation of even small systems. A quantum computer is needed to conduct accurate quantum simulations, which will be introduced in the subsequent chapter.

## IV. QUANTUM TECHNOLOGIES

To simulate nature accurately - quantum physically - classical computers will be overwhelmed no matter how powerful these become in the distant future [30–32]. The inherent complexity that inhabits the quantum scale, including an exponential increase in computational complexity with each additional interaction, can only be dealt with by a special quantum technology - a quantum computer. With the advent of quantum technologies that build upon the most fundamental physical laws of the universe, the question of whether the universe we inhabit and everything in it can be simulated, potentially on a quantum computer, does not seem so obscure anymore. The first quantum revolution, ushered in by the groundbreaking research and discoveries of the great physicists of the early 20$^{th}$ century, not only fertilized many of the exponential technological developments of the last few decades but made them possible [33,34]. The development of lasers has brought us fiber optic communication, laser printers, optical storage media, laser surgery, and photolithography in semiconductor manufacturing, among other things. Atomic clocks gave rise to the global positioning system (GPS), used for navigation and mapping, among other things, and transistors made modern computers possible. These and other technologies of the first quantum revolution are crucial for humanity and the economy. Much of the technology we take for granted in our everyday lives came to light during the first quantum revolution. The first quantum revolution was characterized by the development of technologies that take advantage of quantum effects; however, we have come to understand that these technologies are not exploiting the full potential of quantum physics. Spurred by immense advances in the detection and manipulation of single quantum objects, great strides are now being made in the development and commercialization of applications in quantum technology, such as quantum computing, communications, and sensors, deemed the second quantum revolution, in which the fundamental properties of quantum physics continue to be used. Particles can not only be in two states simultaneously, as is the case with the atoms in an atomic clock. Under certain conditions, two particles at a great distance from each other sense something about the state of the other - they influence each other, which is called entanglement and was already suspect to Albert Einstein. A particle's precise position or state is unknown until a measurement is made. Instead, nature shows us that there are only probabilities of any given outcome, and measuring - looking - changes the situation irrevocably. In comparison, the first quantum revolution was about understanding how the world works on the tiny scales where quantum mechanics reigns. The second is about controlling individual quantum systems, such as single atoms [33,35]. Quantum-mechanical predictions based thereupon are used to measure previously unachieved precision. It is also possible to generate uncrackable codes that cannot be decrypted by any system, thereby forming the basis for a secure communication network [36–38]. In addition, quantum computers promise to solve some currently unsolvable problems, including the simulation of molecules and their interactions for developing drugs against diseases that cannot yet be cured or finding new materials [39,40]. Hybrid computer systems combining classical high-performance computing with quantum computers are already being used today to develop solutions in mobility, finance, energy, aerospace, and many other sectors [41–47]. All these developments are happening right now, and it is remarkable that many challenges are no longer strictly scientific but are now engineering in nature. For example, work is being done on the miniaturization of atomic clocks and the robustness of quantum bits, the information units of a quantum computer. In addition, there are already approaches to amplifying and forwarding quantum communication signals to make internet-based communication more secure than ever before [36–38].

The quantum computer, in which big hopes for simulating physics and chemistry quantum-physically lie, is the quantum technology to be discussed and analyzed in more detail when thinking about simulating the universe. There has been much debate about whether a sufficiently powerful and error-corrected quantum computer may be used to simulate the universe or parts of it. If by "the universe" we are referring to the universe we inhabit, then even with quantum computers using billions of high-quality quantum bits, this will not be possible - both physical and computational constraints discussed in the following chapters prevent us from doing so. A quantum processor or quantum computer is a processor that uses the laws of quantum mechanics. In contrast to the classical computer, it does not work based on electrical but quantum mechanical states. The superposition principle - quantum mechanical coherence – is, for example, analogous to the coherence effects and, secondly, quantum entanglement, corresponding to classical correlation, albeit stronger-than-classical.

Studies and practical implementations already show that using these effects, specific computer science problems such as searching large databases and the factorization of large numbers can be solved more efficiently than with classical computers. In addition, quantum computers would make it possible to significantly reduce the calculation time for many mathematical and physical problems. Before discussing the simulation of parts of the universe utilizing a quantum computer and certain constraints, it is essential to understand the differences between how information is processed in classical and quantum computers. In a classical computer, all information is represented in bits. A bit is physically realized via a transistor, in which an electrical potential is either above or below a certain threshold.

In a quantum computer, too, information is usually represented in binary form. Here, one uses a physical system with two orthogonal base states of a complex two-dimensional space as it occurs in quantum mechanics. In Dirac notation, one basic state is represented by the quantum mechanical state vector $|0\rangle$, the other by the state vector $|1\rangle$. These quantum-mechanical two-level systems can be, for example, the spin vector of an electron pointing either "up" or "down". Other implementations use the energy level in atoms or molecules or the direction of current flow in a toroidal superconductor. Often only two states are chosen from a larger Hilbert space of the physical system, for example, the two lowest energy eigenstates of a trapped ion. Such a quantum mechanical two-state system is called a quantum bit, in short, qubit. A property of quantum mechanical state vectors is that they can be a superposition of other states. A qubit does not have to be either $|0\rangle$ or $|1\rangle$, as is the case for the bits of the classical computer. Instead, the state of a qubit in the complex two-dimensional space mentioned above is given by $|0\rangle := \begin{pmatrix}1\\0\end{pmatrix}; |1\rangle := \begin{pmatrix}0\\1\end{pmatrix}$. A superposition $|\psi\rangle$ is then generally a complex linear combination of these orthonormal basis vectors (Eq. 5), with $c_0, c_1 \in \mathbb{C}$.

$$|\psi\rangle = c_0|0\rangle + c_1|1\rangle = \begin{pmatrix}c_0\\c_1\end{pmatrix} \tag{5}$$

$$|c_0|^2 + |c_1|^2 = 1 \tag{6}$$

As in coherent optics, any superposition states are allowed. The difference between classical and quantum-mechanical computing is analogous to that between incoherent and coherent optics. In the first case, intensities are added; in the second case, the field amplitudes are added directly, as in holography. For normalization, the squared amplitudes sum to unity (Eq. 6), and without loss of generality, c sub 0 can e real and non-negative. The qubit is usually read out by measuring an observable that is diagonal and non-degenerate in its basis $\{|0\rangle, |1\rangle\}$, e.g., $A = |1\rangle\langle 1|$. The probability of obtaining the value 0 as a result of this measurement in the $|\psi\rangle$ state is $P(0) = |\langle 0|\psi\rangle|^2 = |c_0|^2$ and that of the result 1 corresponding to $P(1) = |\langle 1|\psi\rangle|^2 = 1 - P(0) = |c_1|^2$. This probabilistic behavior must not be interpreted so that the qubit is in the state 0 with a certain probability and the state 1 with another probability, while other states are not allowed. Such exclusive behavior could also be achieved with a classical computer that uses a random number generator to decide whether to continue calculating with 0 or 1 when superimposed states occur. In statistical physics, which in contrast to quantum mechanics is incoherent, such exclusive behavior is considered; however, in quantum computing, the coherent superposition of the different basis states, the relative phase between the different components of the superposition, and, in the course of the calculation, the interference between them is crucial. As with the classical computer, several qubits are combined into quantum registers. According to the laws of many-particle quantum mechanics, the state of a qubit register is then a state from a $2^N$-dimensional Hilbert space, the tensor product of the state spaces of the individual qubits. A possible basis of this vector space is the product basis over the basis 0 and 1. For a register of two qubits, one would get the basis 00. The state of the register can consequently be any superposition of these basis states, i.e., it has the form of Eq. 7.

$$|\psi\rangle = \sum_{i_1,\dots,i_N} c_{i_1\dots i_N}|i_1 i_2 \dots i_N\rangle \tag{7}$$

$c_{i_1\dots i_N}$ are arbitrary complex numbers, and $i_1 i_2 \dots i_N \in 0,1$, whereas in classical computers, only the basis states appear. The states of a quantum register cannot always be composed from the states of independent qubits, Eq. 8 showing such an example state.

$$|\psi\rangle = \frac{1}{\sqrt{2}}(|01\rangle + |10\rangle) \tag{8}$$

The state in Eq. 8 and others cannot be decomposed into a product of a state for the first qubit and a state for the second qubit. Such a state is, therefore, also called entangled. Entanglement is one reason why quantum computers can be more efficient than classical computers. Quantum computers can solve certain problems exponentially faster than classical computers: $N$ bits of information are required to represent the state of a classic $N$-bit register. However, the state of the quantum register is a vector from a $2^N$-dimensional vector space, so that $2^N$ complex-valued coefficients are required for its representation. If $N$ is large, the number $2^N$ is much larger than N itself. The principle of superposition is often explained such that a quantum computer could simultaneously store all $2^N$ numbers from 0 to $2^N - 1$ in a quantum register of $N$ qubits, but this notion is misleading. Since a measurement made on the register always selects exactly one of the basis states, it can be shown using the Holevo theorem that the maximum accessible information content of an $N$-qubit register is exactly $N$ bits, exactly like that of a classical $N$-bit register. However, it is correct that the principle of superposition allows parallelism in the calculations, which goes beyond what happens in a classical parallel computer. The main difference to the classical parallel computer is that the quantum parallelism enabled by the superposition principle can only be exploited through interference. For some problems, a greatly reduced running time can be achieved with quantum algorithms compared to classical methods. When it comes to complex computational tasks - and simulating the universe is, without doubt, very computationally intensive and complex - what can be computed with a quantum computer and classical computers is an interesting question. Since the way a quantum computer works is formally defined, the terms known from theoretical computer science, such as computability or complexity class, can also be transferred to a quantum computer. It turns out that the number of computable problems for a quantum computer is no greater than for a classical computer. That is, the Church-Turing thesis also applies to quantum computers. However, there is strong evidence that some problems can be solved exponentially faster with a quantum computer. The quantum computer thus represents a possible counterexample to the extended Church-Turing thesis. A classical computer can simulate a quantum computer since the action of the gates on the quantum register corresponds to matrix-vector multiplication. The classical computer now simply has to carry out all these multiplications to transfer the initial to the final state of the register. The consequence of this ability to simulate is that all problems that can be solved on a quantum computer can also be solved on a classical computer; however, it may take classical computers thousands of years, whereas a quantum computer may take seconds. Conversely, this means that problems like the halting problem cannot be solved even on quantum computers and implies that even a quantum computer is not a counterexample to the Church-Turing thesis. Within the framework of complexity theory, algorithmic problems are assigned to so-called complexity classes. The best-known and most important representatives are the classes P and NP. Here, P denotes those problems whose solution can be calculated deterministically in a polynomial running time to the input length. The problems for which there are solution algorithms that are non-deterministic polynomial lie in NP. For quantum computers, the complexity class BQP was defined, which contains those problems whose running time depends polynomially on the input length and whose error probability is less than $\frac{1}{3}$. Non-determinism allows different possibilities to be tested at the same time. Since current classical computers run deterministically, non-determinism has to be simulated by executing the various possibilities one after the other, which can result in the loss of the polynomiality of the solution strategy. With these results and definitions in mind, it is now time to discuss the potential feasibility of simulating the universe or parts of it.

## V. FATE OF THE UNIVERSE

### A. The beginning of the universe

A simulation of the universe or parts of it requires accurately simulating its evolution from the beginning. In cosmology, the Big Bang that followed cosmic inflation is the starting point of the emergence of matter, space, and time. According to the standard cosmological model, the Big Bang happened about $13.8 \times 10^9$ years ago. "Big Bang" does not refer to an explosion in an existing space but the co-emergence of matter, space, and time from a primordial singularity. This results formally by looking backward in time at the development of the expanding universe up to the point at which the matter and energy densities become infinite. Accordingly, shortly after the Big Bang, the universe's density should have exceeded the Planck density. The general theory of relativity is insufficient to describe this state; however, a yet-to-be-developed theory of quantum gravity is expected to do so. Therefore, in today's physics, there is no generally accepted description of the very early universe, the Big Bang itself, or the time before the Big Bang. Big Bang theories do not describe the Big Bang itself but the early universe's temporal development after the Big Bang, from Planck time (about $10^{-43}$ seconds) after the Big Bang to about 300,000 to 400,000 years later, when stable atoms began to form, and the universe became transparent. The further evolution of the universe is not considered the area of the Big Bang. The Big Bang theories are based on two basic assumptions:

1) The laws of nature are universal, so we can describe the universe using the laws of nature that apply near Earth today. To be able to describe the entire universe in each of its stages of development based on the laws of nature known to us, it is essential to assume that these laws of nature apply universally and constantly, independent of time. No observations of astronomy going back about $13.5 \times 10^9$ years - or paleogeology going back $4 \times 10^9$ years - challenge this assumption. From the assumed constancy and universality of the currently known laws of nature, it follows that we can describe the development of the universe as a whole using the general theory of relativity and the processes taking place there using the standard model of elementary particle physics. In the extreme case of high matter density and, at the same time, high spacetime curvature, the general theory of relativity and the quantum field theories on which the Standard Model is based are required for the description. However, the unification encounters fundamental difficulties such that, at present, the first few microseconds of the universe's history cannot be consistently described.
2) The universe looks the same at any place (but not all times) in all directions for considerable distances. The assumption of spatial homogeneity is called the Copernican principle and is extended to the cosmological principle by the assumption of isotropy. The cosmological principle states that the universe looks the same simultaneously at every point in space and in all directions for large distances, which is called spatial homogeneity. The assumption that it looks the same in every direction is called spatial isotropy. A look at the starry sky with the naked eye shows that the universe in the vicinity of the Earth is not homogeneous and isotropic because the distribution of the stars is irregular. On a larger scale, the stars form galaxies, partially forming galaxy clusters distributed in a honeycomb structure composed of filaments and voids. On an even larger scale, however, no structure is recognizable. This and the high degree of isotropy of the cosmic background radiation justify the cosmological principle's description of the universe as a whole. If one applies the cosmological principle to the general theory of relativity, Einstein's field equations are simplified to the Friedmann equations, which describe a homogeneous, isotropic universe. To solve the equations, one starts with the universe's current state and traces the development backward in time. The exact solution depends, in particular, on the measured values of the Hubble constant and various density parameters that describe the mass and energy content of the universe. One then finds that the universe used to be smaller. At the same time, it was hotter and denser. Formally, the solution leads to a point in time when the value of the scale factor disappears, i.e., the universe has no expansion, and the temperature and density become infinitely large. This point in time is known as the "Big Bang". It is a formal singularity of the solution of the Friedmann equations. However, this does not make any statement about the physical reality of such an initial singularity since the equations of classical physics only have a limited range of validity and are no longer applicable when quantum effects play a role, as is assumed in the very early, hot and dense universe. A theory of quantum gravity is required to describe the universe's evolution at very early times.

*B. The future of the universe*

he brightest stars determine the brightness of galaxies. In our galaxy, there are 100 billion stars, 90 percent of which are smaller than the Sun, and 9 percent of those are about as large as our Sun and up to about 2.5 times more massive. Only one percent is much larger than the Sun. However, this one percent of the stars determine the galaxy's brightness. The bigger a star is, the more wasteful it is with its nuclear fuel. The brightest stars only live tens of millions of years and explode at the end of their lives. In doing so, enrich the interstellar gas with heavy elements. Besides, the shock waves of the explosion compress the gas so that new stars can emerge. Due to this wasteful handling of nuclear fuel, however, there will, at some point, be no more gas available for new stars to form. Our Sun already contains 2 percent heavy elements as the proportion continues to increase, so in a period of no more than a thousand billion years from now, gas will no longer be available to form new stars. Our galaxy will then shine weakly by the smaller stars that live much longer than the Sun. The smallest of them will go out only after $10^{12} - 10^{13}$ years. However, these already glow so faintly that to our eyes - if there are still humans in their current form - the sky would look almost starless, as we would only recognize such faint stars nearby. In about $10^{13}$ years, the universe will slowly become dark. The universe then only consists of slowly dying white dwarfs, planets, pulsars - the 20-kilometer cores of giant stars with a density as in an atomic nucleus -, and black holes. Even though it is dark, the gravitational forces of the stars still exist. Stars are usually far apart. The Earth is eight light minutes away from the Sun, and the next star is 4.3 light-years. So, a very close encounter with another star, though unlikely, will be enough to throw Earth out of its orbit around the long burned-out Sun. In that scenario, the Earth wanders alone around the Milky Way. That an encounter with a massive partner is so close that it also affects the Sun is even more unlikely, but even such an event should have occurred once in $10^{19}$ years. The Sun either falls closer to the galactic center or is expelled from our galaxy. Stars not forced out of our galaxy will fall victim to Sagittarius $A^*$, the black hole in the center of it. Currently, it has a mass of $4.31 \pm 0.38 \times 10^6$ solar masses, but after $10^{24}$ years, all the stars remaining in the system have likely been swallowed. According to today's model of the universe's origin, we assume that it originated from a point, a singularity of all mass, and since then, it has continued to expand. Whether this will continue depends on how much matter is in the universe. The matter slows down the expansion by gravitational attraction. If there is enough matter in the universe, the expansion can eventually come to a halt and then turn around so that the universe ends in a singularity. The universe will continue expanding if the matter does not suffice for that scenario. The amount of material required for this is called critical density. The matter we can observe (stars, gas clouds) is not enough for this, but we suspect that there may still be matter that we cannot see, the so-called dark matter. This could be elementary particles, black holes, or prevented stars where the mass was not sufficient to ignite the nuclear fuel. There are indications of substantial amounts of dark matter since the mass in galaxy clusters is not enough to hold them together. According to the theory of relativity, space and time interact with matter and energy. The gravitation of matter in the universe slows down cosmic expansion. In the distant past, the expansion rate should have been greater. If the mean density of matter in the universe is above a specific "critical" value, then the expansion should even come to a halt at some point and turn into a contraction. This critical density is around two to three hydrogen atoms per cubic meter, corresponding approximately to the mass of a grain of sand distributed over the volume of the Earth. However, the observations show that the total mass of the visible matter - gas, stars, and dust – is at most enough to accommodate one percent of the critical density. As of our current understanding, the main mass of the universe consists of dark matter. Dark, because it only makes itself felt through the force of gravity in the movements of stars and galaxies, but it's not observable. However, the density only comes to about 30 percent of the critical limit. Research into the galaxy clusters now speaks for this: their spatial distribution and dynamics, their temporal development based on observations and theoretical models, and the extent of gravitational lens effects - light deflection of distant galaxies by foreground objects. So, the universe's average density is not enough to stop its expansion. Even so, theorists have long favored a critical density universe because this is the simplest solution that best fits the cosmic inflation hypothesis. Cosmic inflation describes that the universe should have expanded exponentially in the first fraction of a second after the Big Bang when it was even tinier than an atom - by a factor of $10^{30}$. The pattern of tiny temperature fluctuations in the cosmic background radiation also speaks for such a universe with critical density. It is the reverberation of the Big Bang and contains subtle information about the entire properties of the universe. The density $\Omega_{tot}$ is given by Eq. 9.

$$\Omega_{tot} = \frac{\rho}{\rho_c} \qquad (9)$$

where $\rho$ is the mean density, and $\rho_c$ the critical density, which in turn is defined in Eq. 10.

$$\rho_c = \frac{3H_0^2}{8\pi G} \approx 8.5 \cdot 10^{-27} \frac{kg}{m^3} \qquad (10)$$

where $H_0$ is the current Hubble parameter giving *the universe's* rate of expansion, and $G$ is the gravitational constant. Whether the mass is sufficient to stop the universe's expansion is unknown today. The observable mass accounts for only $10 - 20$ percent of the critical density.

- Critical density > 1: In this case, the universe's expansion will reverse, and the universe will collapse in the distant future. For an observer, the reversal of expansion would not be apparent at first. Only when there are only a billion years left before collapse do observers notice that the sky is getting lighter again. More and more galaxies will appear in the sky as the distances between the galaxies get smaller. Around this time, the galaxy clusters will also merge. About $100 \times 10^6$ years before the collapse, the galaxies will merge, and stars will only be found in a gaseous cloud. One million years before the collapse, the temperature in space will rise to room temperature. One hundred thousand years before the collapse, the night sky will be as bright as the Sun's surface. The temperature in the universe will rise to such an extent that planets become liquid lumps. The further the matter is smushed together, the faster the black holes grow. Besides, neutron stars and dwarf stars can form new black holes by absorbing matter. Ten years before the collapse, even the black holes merge. The temperature in the universe is now $10 \times 10^6$ degrees. Finally, the universe is only a black hole, so it does not matter if it comes to collapse since, in a black hole, both time and space no longer exist. Otherwise, everything now runs backward, as in the Big Bang: new energy spontaneously forms new particles, the basic forces of nature separate, and everything ends in a great singularity.
- Critical density = 0: The most boring case. If the density of matter has precisely the critical value, the expansion rate will increasingly approach zero in an infinite time. The conditions in this universe are similar to those in a universe with an omega of less than 1. However, an electron and a positron will circle each other at astronomical distances. These come closer and closer, eventually annihilating each other and leaving behind photons. The end is a shapeless desert of radiation and particles that is lost in eternity. If nothing happens, then the concept of time no longer makes sense, and time ceases to exist.
- Critical density < 1: After $10^{14}$ years, hydrogen fusion to higher elements in the stars comes to a standstill. They slowly go out one by one. The universe now consists only of planets, dwarf stars, neutron stars, and black holes. In $10^{17}$ years, the radiation of light from residual gas will also slow down and hit the remaining stars. In $10^{26}$ years, galaxies emit gravitational radiation. The remaining stars are slowly spiraling into the galaxy center. If current theories about the universe's origin apply, the proton would have to split with halftime of $10^{32} - 10^{36}$ years. Some theories that predict a decay time of fewer than $10^{34}$ years can already be excluded, as no proton decay could be observed so far. Therefore, it is impossible to say that the theories predicting a longer decay time are correct. If true, half of the remaining matter in the universe will have decomposed into positrons after $10^{36}$ years. Accordingly, all protons in space would have decayed after $10^{39}$ years at the latest. That would also be the end of all atoms, molecules, planets, and other celestial bodies since the atomic nuclei partially consist of protons. The universe would now only consist of light, electrons, positrons, and black holes. After $10^{64} - 10^{67}$ years, the black holes begin to evaporate slowly. The last and formerly biggest ones will end in an explosion after $10^{100}$ years. Whether black holes actually evaporate is, as of now, only a hypothesis. If the proton does not decay, after $10^{1000}$ years, the dwarf stars will become neutron stars. After $10^{10^{26}} - 10^{10^{74}}$ years, the neutron stars will have evolved into black holes, which will then evaporate. After this time, the universe will consist of elementary particles only.

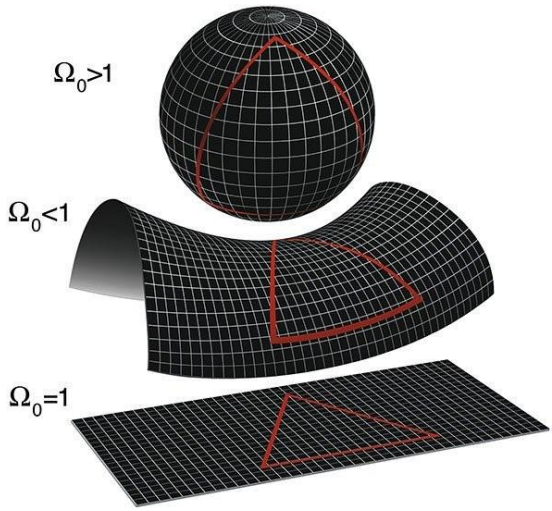

Fig. 1. Curvature of the universe depending on the density parameter [48]

Independent of the possible futures of the universe, according to human standards, it has taken a long time to reach the current state of the universe, and the evolution will continue for a long time before its thermodynamic death or collapse happens. Therefore, a simulation of the universe or parts of it running in real-time will be no good if we expect to learn whether the simulation behaves like the real universe. One would assume that, like in today's computer simulations, time can be accelerated to gain insights more quickly. Still, it turns out that trying to simulate the universe exactly imposes some constraints on the simulation time, as is discussed in the following chapters.

## VI. SIMULATION FROM WITHIN AND FROM WITHOUT

Before discussing the simulation of our universe, we must distinguish between a simulation of our current universe and an arbitrary universe. The question of whether it would be possible to predict the future given an exact simulation of the universe we inhabit is not only interesting to ask but important to answer, as it leads us towards the ultimate limit of computability, both in terms of physical law and complexity. We distinguish between the simulation of *the universe* and the simulation of *a universe*, the former referring to the exact simulation of the universe we inhabit (our universe), the latter referring to either part of *the universe* or an unspecific universe, which may or may not be based on the same physical laws as *the universe*. A couple of assumptions are made concerning our understanding of the evolution of *the universe* and everything in it, including astronomical objects such as planets, stars, nebulae, galaxies, and clusters, as well as dark energy, dark matter, the interstellar medium, the expansion of space-time, and the geometry of *the universe*, and the universal constants - the speed of light in vacuum, Planck's constant, Boltzmann's constant, and the gravitational constant, to name a few. For our today's knowledge of *the universe* is far from complete, for the sake of the subsequently presented arguments, some assumptions are made:

1) We (or an external programmer) completely understand the physical laws governing evolution and *the universe's* composition down to level X. This includes a complete understanding of *the universe's* cosmology, including all astronomical objects, the beginnings of the Big Bang and inflation to all matter - including dark matter -, dark energy, and all laws at the respective scale, such as relativity and quantum physics. A complete understanding of all physical laws allows us (or an external programmer) to appropriately model *the universe's* evolution and simulate parts of it. It also allows life to emerge in the simulation. This assumption is not true today.

2) The laws of quantum physics are the most fundamental physical laws down to level X. The constraints quantum physics imposes on certainty govern how accurately we can measure in our universe. The Planck units are the accessible limits of time, space, mass, and temperature beyond which no physical law has meaning. There are scientifically sound experiments and strong indications that this is true, but many open questions remain, for example, how to unite quantum physics with general relativity. Thus, this assumption is partially true today.

3) An external programmer may decide to simulate *a universe* utilizing the same physical laws prevailing in their universe or base the simulation on different physical laws. If we act as an external programmer intending to find indications as to whether we participate in a simulation chain, it seems reasonable to base the simulation on the same physical laws prevailing in *the universe*. An external programmer to *the universe* also can't know whether they participate in a simulation chain, and the assumption we make is that an external programmer would also want to find indications for whether it is true or not and would base a simulation of *a universe* on the same physical laws they observe in their universe. In fact, since we are pondering about simulating *a universe*, the question of whether *the (our) universe* is based on the same physical laws as a potential external programmer's universe becomes irrelevant: if we act as external programmers and base our simulation on the same physical laws as our universe, and if a simulation chain emerges in which subsequent external programmers base their simulations on the same physical laws, too, then the actions of these external programmers are candidate observations to look for in *the universe*, no matter if the latter is based on the same physical laws of a potential external programmer's universe.

Level X refers to the deepest physical level of structure of matter and the physical laws governing this level. As of today, we do not know with certainty what the deepest structure of matter is, so for now, we call it X. One of the hypotheses of string theory is that quarks - the constituents of hadrons, including protons and neutrons - and electrons are made up of even smaller vibrating loops of energy called strings and that these are the most fundamental elements all matter is made of. It will be essential to understand if X is also the level of granularity a "sufficiently good" simulation needs to be based upon, as, for example, a macroscopic engine's workings do not seem to be influenced too much by individual quark behavior.

Also, a distinction needs to be made between simulation and emulation. On the outside, an emulation of *the universe* behaves exactly as the real universe, but the emulation's internal states do not need to be identical to the real universe's. For example, an emulation would not have to consider the same physical laws and behavior on unobserved scales, i.e., down to level X, as long as it behaves exactly as *the universe* on the desired observed scales. On the other hand, a simulation of *the universe* represents in its internal state all physical laws and states of the simulated constituents precisely as they are outside the simulation. Both the simulation and the emulation of *the universe* may produce convincing results. However, a simulation forms the basis for further discussion, as the intention is not only to mimic behavior but to reproduce parts of *the universe*, making the simulation physically indistinguishable from the real universe, both internally and externally.

*A. Simulation of the universe*

The feasibility of simulating *the universe*, specifically, the exact simulation of *the universe* we inhabit in full size and complexity, which would allow us to predict future events and simulate arbitrary states backward in time exactly, can logically be ruled out, even under the above assumptions. Several constraints on computability prevent the creation of such a simulation. If the intention is to simulate *the universe* exactly, the simulation must also include the computer used for the simulation, which we call the "simulation from within". The "simulation from without", on the other hand, is one in which the internal state of the computer used for the simulation of *the universe* is decoupled from *the universe's* state.

- Every simulation takes discrete steps in time and will predict one time step after the other until the desired prediction time has been reached. Let us assume the simulation starts at time $t_0$ with the expectation of simulating $t_1$ to predict the state of *the universe* at $t_1$, and the computing time $t_c$ to perform this is smaller than $t_1$ (see Eq. 11).

$$t_0 = t_1 - t_c \qquad (11)$$

If the computer predicted the state of *the universe* at $t_1$ in $t_c$, it also predicted its internal state at $t_1$. If the computer is asked at $t_c$ to predict the state of *the universe* at $t_2$, it will base its predictions on the state of *the universe*, including its internal configuration at $t_1$. However, $t_c < t_1$, so the computer would start to change its internal configuration to predict *the universe*'s state at $t_2$ before $t_1$ has been reached and to its original predictions of its internal state at $t_1$ are not correct when $t_1$ is reached. No matter how small the time steps are, as long as $t_c$ is smaller than $t_1$, this problem arises, leading to the conclusion that a computer within *the universe* can't be used to simulate *the universe*; it can only simulate *a universe*, according to our previously made definitions. The core of this argument is that a simulation of *the universe* can't run faster than *the universe's* real-time evolution and that *the universe's* future can't be predicted if the prediction should be exact. As the computer will always predict its own incorrect internal state as long as $t_c < t_1$. If $t_c = t_1$ and no other constraints are given, the simulation would produce correct results but run in real-time. From here on, we call this the **computational predictability constraint**.

- The first reason for the incomputability of the simulation from within is that - assuming it would be possible to run a simulation from within, including a correct representation of the internal state of the computer at any given point in time - every simulated computer would also have to simulate *the universe* including itself, which is a recursion. The recursion is not only computationally expensive, but it also results in general incomputability of the simulation, as the available computational resources, such as memory and processing units, have to be used to compute a set of simulations $S = \{s_1, \ldots, s_n\}$ where $\lim_{n \to \infty}$ . Instead of the computational resources needed for simulating *the universe* within, each simulated computer needs the computational resources to simulate itself and *the universe*, meaning that the computer in $s_1$ is required to simulate *the universe* and itself from within, and infinitely many times from without, which is incomputable and thus impossible. Either the computer in $s_1$ would run out of resources at some point, or the computational resources in $s_n$ would not suffice to conduct another simulation $s_{n+1}$. From here on, this argument will be referred to as the **first computability constraint**.

- Yet another reason for the incomputability of the simulation from within is *the universe's* complexity. We are used to computers providing us access to virtual worlds via virtual reality devices or screens, and the virtual worlds of video games and the metaverse have become increasingly complex even though the information processing happens on purely classical computers that do not even have to be enormously powerful. Algorithms can be used to generate environments and mimic infinity randomly and dynamically. This is misleading though, when it comes to the simulation of *the universe* from within, as a computer running such a simulation would have to represent all particles of *the universe*, which we currently assume $10^{78} - 10^{82}$, in its internal state. A fundamental question is whether the encoding object can be of simpler nature than the encoded object. Today's most advanced quantum computers use quantum objects to encode states of other systems but are limited to binary states. For example, such a quantum computer may use the energy levels of atoms to encode binary states - the ground state to encode 0 and an excited state to encode 1. A bit, no matter if it's a quantum bit or a classical bit, is the smallest unit of information, and a physical system able to hold the information of one bit is the most fundamental system for storing information. How many bits are needed to completely describe an atom? An atom itself is a lot more complex than a two-state system. For example, electrons are found in probability clouds around atomic nuclei, called orbitals (Fig. 2).

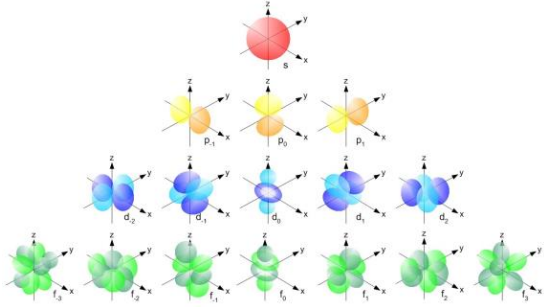

Fig. 2. Some electronic orbitals. An orbital is a probability cloud determining the position of an electron within an atom [49].

Each electron in an atom is described by four distinct quantum numbers, providing information on the energy level or the distance of the electron from the nucleus, the shape of the orbitals, the number and orientation of the orbitals, and the direction of the electronic spin. These four quantum numbers contain complete information about the trajectories and the movement of each electron within an atom. All quantum numbers of all electrons combined are described by a wave function corresponding to the Schrödinger equation. The Schrödinger equation is simple to understand, but even the most powerful quantum computers of the future will need many more atoms to encode the states of a system than the system in simulation consists of. The hydrogen atom, for example, consists of an electron and a proton. Both particles move around a common center of mass, and this internal motion is equivalent to the motion of a single particle with reduced mass. $r$ is the vector specifying the location of the reduced particle's electron relative to its proton's position. The orientation of the vector pointing from the proton to the electron gives the direction of $r$, and its length is the distance between the two. In today's simulations, approximations are made, for example, that the reduced mass is equal to the electron's mass and that the proton is located at the center of the mass. Eq. 12 is the Schrödinger equation for the hydrogen atom, with $E$ giving the system's energy.

$$\widehat{H}(r,\theta,\phi)\psi(r,\theta,\phi) = E\psi(r,\theta,\phi) \quad (12)$$

The time evolution of a quantum state is unitary, and a unitary transformation can be seen as a rotation in Hilbert space. Evolution happens via a special self-adjunct operator called the Hamiltonian $\widehat{H}$ of the system. The equation is given in spherical coordinates, with $r$ determining the radius, and $0 \leq \phi < 2\pi$ being the azimuth and $0 \leq \theta \leq \pi$ the polar angle. The time-independent Schrödinger equation for an electron around a proton in spherical coordinates is then given by Eq. 13.

$$\begin{aligned}\left\{-\frac{\hbar^2}{2\mu r^2}\left[\frac{\partial}{\partial r}\left(r^2\frac{\partial}{\partial r}\right)\right.\right.&\\ +\frac{1}{\sin(\theta)}\frac{\partial}{\partial\theta}\left(\sin(\theta)\frac{\partial}{\partial\theta}\right)&\\ \left.+\frac{1}{\sin^2(\theta)}\frac{\partial^2}{\partial\phi^2}\right]&\\ \left.-\frac{e^2}{4\pi\epsilon_0 r}\right\}\psi(r,\theta,\phi)&\\ =E\psi(r,\theta,\phi)&\end{aligned} \quad (13)$$

In Eq. 13, $\hbar$ is the reduced Planck's constant, the term in the squared brackets describes the kinetic energy, and the term subtracted from the kinetic energy is the Coulomb potential energy. $\psi(r,\theta,\phi)$ is the wave function of the particle, $\epsilon_0$ the permittivity of free space, and $\mu$ is the two-body reduced mass of the hydrogen nucleus of mass $m_p$ and the electron of mass $m_q$ (Eq. 14).

$$\mu = \frac{m_q m_p}{m_q + m_p} \quad (14)$$

It is possible to separate the variables in Eq. 13 since the angular momentum operator does not involve the radial variable $r$, which can be done utilizing a product wave function. A good choice is Eq. 15, because the eigenfunctions of the angular momentum operator are spherical harmonic functions $Y(\theta,\phi)$ [50,51].

$$\psi(r,\theta,\phi) = R(r)Y(\theta,\phi) \quad (15)$$

The radial function $R(r)$ describes the distance of the electron from the proton, and $Y(\theta,\phi)$ provide information about the position in the orbital, and a solution to both with $E_n$ depends only on the primary quantum number $n$ (Eq. 16).

$$E_n = \frac{m_e e^4}{8e_0^2 h^2 n^2} \quad (16)$$

The atom's wave functions $\psi(r,\theta,\phi)$ are the atomic orbitals discussed before, and each of the orbitals describes one electron in an atom. Considering only distinct orbitals in an atom, say from 1s up to 5g, the system used for encoding these would have to be able to use 6.8 bits, resulting from the base 2 logarithm of 110 (Eq. 17). According to quantum theory, there is an infinite number of combinations of these four quantum numbers per atom, which would require the encoding system to use an infinite number of encoding systems. Also, as of today, the Schrödinger equation can only be solved for the simplest systems, such as hydrogen-like atoms. This limitation is not a constraint by computers, as also quantum computers will not be able to solve the Schrödinger equation analytically for big systems. However, quantum computers will provide a speedup when solving it numerically. Therefore, the assumption seems reasonable that to encode the full complexity of a quantum system, an identical quantum system is needed.

$$\log_2(110) = 6.8 \qquad (17)$$

To achieve this feat, the computer, if it is governed by the same physical laws as the simulation, would at least have to consist of the same number of particles and be the size of the universe. "At least", because the objects used by the computer to encode objects in the universe must be of equal representational power, which is impossible and therefore proves the simulation's incomputability from within. It has been found that there is a strong connection between the ultimate information capacity of devices and the generalized second law of thermodynamics (GSL) [52]. It was shown that the total area of event horizons in black holes in certain black hole events, such as mergers, never decreases [53]. GSL states that any time matter falls into a black hole, the increase in black hole entropy (over)compensates for the loss of the matter's entropy, preventing both the entropy outside a black hole and the black hole entropy from decreasing. It was further theoretically determined that the proportionality constant between black hole entropy and the area of the event horizon of the black hole is one quarter of the event horizon's area in Planck areas (Eq. 18) [52,54], where one Planck area is the Planck length of $10^{-33}$ cm squared, which would result in an information-theoretical entropy [55] of $10^{66}$ bits for a black hole with a diameter of one centimeter. Entropy, as defined by information theory (Eq. 20), differs from thermodynamic entropy (Eq. 19) referred to in this publication before, specifically in how it is calculated. In Eqs. 18, 19, 20, S is the thermodynamic entropy, $H$ is the information theoretical entropy, $k_B$ is the Boltzmann constant, $l_p^2$ is the Planck length, $W$ is the thermodynamic probability defining the number of alternative microscopic arrangements corresponding to the same macroscopic state, and $X$ is a discrete random variable taking values in $\chi$.

$$S_{BH} = \frac{k_B A}{4 l_p^2} \qquad (18)$$

$$S = k_B \ln W \qquad (19)$$

$$\sum_{x \in \chi} p(x) \log_b p(x) \qquad (20)$$

The GSL allows for setting bounds on the information storage capacity of any isolated system, and groundbreaking works on the holographic bound [56–59] brought to light that the maximum possible entropy of a system depends on the boundary area, or surface, of that system instead of its volume. For a certain volume of space, the holographic bound limits how much entropy in energy and matter can be contained. The surprising connection between entropy and the surface of a black hole, as well as the enormous amount of information that can theoretically be stored on the surface of a small black hole, shows that even spatially small theoretical information processing devices can store huge amounts of data, such as $10^{66}$ bits for a black hole with a diameter of one centimeter. However, if we were to simulate *the universe,* including all its objects, certainly also black holes governing the behavior of galaxies and, indirectly, structures beyond would have to be simulated. The argument that a very powerful but small information processing device could be used to simulate *the universe does* not hold. However, these results provide strong indications that a device simulating *a universe* not only need not be the size of *a universe* but can be much smaller than currently conceivable based on our most advanced information processing devices. From here on, this argument will be referred to as the **second computability constraint**.

- Lastly, if quantum physics comprises the most fundamental physical laws, which we currently understand to be the case, it imposes certain limits on how accurate our knowledge about particles can be. Even though quantum physics comprises the fundamental theories for describing *the universe*, the conceptual structure of these theories differs profoundly from that of classical physics - the physics we experience every day when interacting with parts of *the universe*. The statements of quantum physics about *the universe* are statements about the results of measurements. In contrast to classical physics, we can make probabilistic statements in each case; for example, one can only predict the value distribution when measuring an ensemble of similar systems. Heisenberg's uncertainty principle (Eq. 21) results from the fact that a physical system is quantum-physically described with the help of a wave function.

$$\Delta x \times \Delta p \approx h \qquad (21)$$

The quantum-physical uncertainty arises in all wave-like systems because of the matter-wave nature of all quantum objects. For example, $\Delta x$ of a photon's location depends on the wavelength of the light under consideration. On the other hand, the deflection of the light quantum acts like an impact on the particle, whereby the body's momentum experiences an indeterminacy of $\Delta p$, known as Compton scattering. As a basic lower limit for these uncertainties, Heisenberg used the de Broglie relationship [60] to estimate that the product of $\Delta x$ and $\Delta p$ cannot be smaller than that for the natural constant characteristic of quantum physics, Planck's constant $h$. Heisenberg formulated this fundamental limit of measurability in the statement given in Eq. 21. While in classical mechanics, location and momentum are simple quantities that can, in principle, be measured precisely, in quantum mechanics, their distributions result as the square of the absolute value of the wave function or its Fourier transform. That is, they cannot be determined independently of one another. Since the distributions of position and momentum depend on the system's wave function, the measurements' standard deviations also depend on each other. The more precisely one wants to determine the location of a particle in the usual quantum mechanical description, the greater the imprecision of the momentum - and vice versa. If quantum physics is the most fundamental set of physical laws - the machine language of *the universe*, so to speak, then the uncertainty relation prevents us from making an accurate simulation of *the universe* from within and without. Additionally, entanglement also prevents us from simulating *the universe* from within, as there is no way of disentangling the particles making up the computer used for the simulation from *the universe*'s particles. Entanglement in quantum physics is when a composite physical system, for example, a system consisting of several particles viewed as a whole, assumes a well-defined state without being able to assign a well-defined state to each subsystem. This phenomenon cannot exist in the field of classical physics. There, composite systems are always separable. Each subsystem has a specific state at all times, determining its respective behavior, with the totality of the states of the individual subsystems and their interaction fully explaining the behavior of the overall system. In a quantum-physically entangled state of the system, on the other hand, the subsystems assume several of their possible states at once, with each of these subsystem states being assigned a different state of the other subsystems. To explain the overall system's behavior correctly, one must consider all these coexisting possibilities together. Nevertheless, when a measurement is carried out on each subsystem, only one of these possibilities is observed, with the probability of measuring that particular result, all results being determined by a probability distribution. Measurement results from several entangled subsystems are correlated with one another. That is, depending on the measurement result from one subsystem, there is a different probability distribution for the possible measurement results from the other subsystems. Entangled states are common; an entangled state arises whenever two subsystems interact with each other, for example, collide with each other, with different but coordinated possibilities for how they continue to behave, such as in which direction they continue to propagate after the collision. According to quantum physics, all of these possibilities come with a certain probability by which they have to be represented in a correspondingly coordinated manner in the state of the overall system up to the moment of the quantum mechanical measurement. Entanglement is destroyed as soon as one of the subsystems is fixed to one of its states. Then another subsystem, linked to the first subsystem by the entanglement, immediately transitions to the state assigned to the state of the first subsystem determined by observation. The state of the overall system then no longer shows any entanglement because both subsystems are now in their specific state. Many experiments have proved the correlations caused by entanglement. These correlations are independent of the distance between the locations at which the measurements are taken on the subsystems and the time interval between the measurements. The same is true if the measurements are so far apart and are carried out so quickly, one after the other, or even simultaneously, that the measurement result on one particle cannot have influenced the state of the other in any physical way. In certain experiments, the correlations are so strong that, in principle, they cannot be explained by any theory that, like classical physics, is based on the physical principle of local realism. Local realism states that each subsystem always has a well-defined state on which another spatially remote subsystem can only act at the speed of light. According to Bell's theorem, it is also ruled out that such a local-realistic theory with hypothetical additional hidden variables could describe the phenomenon of quantum correlation. There is still scientific debate about whether the incompatibility of quantum physics with local hidden variable theories and the probabilistic nature of particles are sufficient to rule out a deterministic universe and whether this would allow for free will [61–63]. Assuming the perspective of an internal or external programmer in the sense of this publication, in order to make an accurate prediction about future states of *a universe* (other constraints left aside), the programmer would want to have a perfect measurement device - a device that is completely isolated from its environment (and so not influenced by its environment) and lets them probe quantum states. In a simulation, a particle needs to exist in full complexity in the internal state of the computer simulating it. Computers with such capabilities do not exist today today's quantum computers are not capable of simulating complex quantum physics or -chemistry. Now, assuming such a computer and such a measurement device exist, a programmer has several options:

- They could run two identical simulations, $s_1$ and $s_2$, in parallel, and once both simulations reached a certain time $t_x$, they could speed up $s_2$ such that it reaches, say $t_{x+2}$ before $s_1$ gets there, and peak into $s_2$ to learn the future state $t_{x+2}$ of $s_1$. However, the programmer would be surprised to learn that already at $t_x$, *the universes* are not identical anymore, as every quantum-physical interaction between particles is probabilistic. Even if the initial configuration of $s_1$ and $s_2$ is identical - which can conceivably be implemented - the evolution of both may be similar but not identical. For example, the result of chemical

- reactions depends on quantum effects, such as electron spin. Moreover, the collapse of systems' wave functions and entanglement and disentanglement between particles happens constantly, and the outcome is not deterministic. Two identical interactions in $s_1$ and $s_2$ may yield different results. Over the course of countless interactions, this will result in distinguishable/different macroscopic states.
- The programmer may just run one simulation $s_1$ and decide to take a "snapshot" of it at $t_x$, and before it reaches $t_{x+2}$ feed the snapshot into another computer for predicting the simulation's state at $t_{x+2}$. It is not possible to create a snapshot incorporating the complete quantum state of all particles at $t_x$, as in quantum physics, the no-cloning [64–66] theorem prevents the programmer from doing so. In short, it proves the impossibility of creating an independent and identical copy of an arbitrary unknown quantum state.
- The programmer could probe every quantum state of the simulation using their perfect measurement device at $t_x$, obtain classical results, and use these to create new quantum states in another computer for predicting the simulation's state at $t_{x+2}$. Here, the challenge is that every measurement device, no matter how good it is, will have to interact with a quantum state, resulting in a change of state of that very quantum state before it collapses (a measurement device always induces collapse of the information-rich but inaccessible superposition of a state onto one definite state). One example is the interaction of a photon with an electron to probe its spin. Moreover, it is impossible to completely isolate a quantum system from its environment [67]. So also, a measurement device made up of many quantum systems cannot be isolated from its environment or its own irrelevant physical properties. Consequently, even if a measurement device gets as good as it can, it is impossible to predict future states using this approach.
- The programmer may pursue a simulation using a computer that simulates quantum states without using quantum effects for computation, in line with the thinking about simulated quantum computers [41, 68]. As the number of quantum states, even for single particles, is infinite, any such computer can only produce approximate simulations, and the computational power needed to simulate even simple systems would increase with the degree of accuracy. The **third computability constraint** outlined here would be fulfilled even earlier.

Whether *the universe* is non-deterministic is subject to intense debate, and arguments for determinism include superdeterminism [61,69] and deterministic quantum mechanics [70]. Deterministic quantum mechanics argues, among other things, for ontological bases, in which the Schrödinger equation sends basis states into other basis states at sufficiently dense moments in time, and that the superimposed states used in quantum mechanics cannot be experimentally produced [70], but are written as such because we lack the knowledge to describe the exact ontological state. This is not yet supported by theory or experiment, as Ontological bases have not yet been found.

Superdeterminism is a hypothetical class of theories that escape Bell's theorem because they are fully deterministic. Bell's theorem assumes that the types of measurements made at each detector can be chosen independently and the hidden variable being measured. For Bell's inequality argument to continue, it must be possible to talk about the experiment's outcome if different choices had been made. In a deterministic theory, the measurements that the experimenters select at each detector are predetermined by the laws of physics. It can be argued that it is wrong to talk about what would have happened if different measures had been chosen: a different choice of measurement was not physically possible. Because the measurements selected are determined in advance, the results at one detector can be affected by the type of measurement made at another detector without the need to transmit the information faster than the speed of light. There is a way to escape the conclusion of superluminous speeds and remote actions through entanglement, but it implies absolute determinism in *the universe*, the total absence of a free will. If we assume that the world is superdeterministic, meaning that not only the inanimate but also our very behavior is predetermined, including our belief that we are free to choose to have one experience over another, then also the decision of what set of measurements an experimenter makes is predetermined, and all difficulties disappear. Following this argument, there is no need for superluminous communication. As for any signal, it is futile to tell particle A what measurement was made on particle B since *the universe*, including particle A, already "knows" what that measurement and its result will be. Bell himself argued that even if deterministic random number generators select the measurements, the choices can be considered effectively free for the object in question since many minimal effects modify the machine's choices. The hidden variable is unlikely to be sensitive to the same small influences as the random number generator [71]. Also, experimental evidence contradicts superdeterminism, as thus far, all Bell tests have found that the hypothesis of local hidden variables is inconsistent with how physical systems behave [72]. Lastly, from a purely logical perspective, superdeterminism is a circular argument as its core assumption is that *the universe* is deterministic; hence, quantum physics must be deterministic.

Also, since there is no way to completely isolate the computer's particles from the rest of *the universe's* particles in the simulation within, the computer's changing internal states will always influence the state of *the universe*, thus rendering an exact simulation of *the universe* incomputable.

The presented arguments are strong indications for a quantum physical universe being non-deterministic and unpredictable. From here on, we call this the **physical predictability constraint**.

The arguments against computability and predictability show clearly why a simulation of *the universe* from within is impossible and will not be possible in the distant future, no matter how powerful computing devices become. Different arguments hold for the simulation of a universe.

*B. Simulation of a universe*

In this text, simulating *a universe* refers to the simulation of parts of *the universe* or some universe with different physical laws. The latter would not be an emulation, as the goal is not to mimic *the universe's* behavior. The simulation of *a universe* includes both the simulation of *a universe* from within and the simulation of *a universe* from without, whereby the latter also includes the simulation of *the universe* from without as for an external programmer, *the universe* is *a universe*, and an external programmer is constrained by the same computational limitations defined previously when trying to simulate their universe. This also holds if we appear as the external programmers to intelligent (for the lack of a better word) beings in *a universe*. In terms of computability, the simulation of *a universe* is certainly possible, and there are various arguments supporting the hypothesis that we ourselves exist in a simulation of *a universe*. We distinguish between the simulation of *a universe* from within and from without, whereby the simulation from within is the exact simulation of parts of *the universe* by an internal programmer, for example, us simulating a galaxy of *the universe*. A simulation from without is an external programmer simulating *a universe* from their within. *The universe* could, for example, be such a simulation. The computer, in that case, can be much smaller and less complex than it would have to be when simulating *the universe*, and the smallest logical units of information qubits, for example -, may be made up of multiple physical particles or systems.

- The simulation of *the universe* from within by an internal programmer cannot contain a simulation of the computer running the simulation in case the $t_c < t_n$, where $t_n$ marks an arbitrary time step in the simulation because of the argument of unpredictability introduced before. However, the simulation of *a universe* excluding the computer used for running the simulation is no contradiction, other constraints set aside. *A universe* simulated from within or without can be smaller than *the universe* of the internal or external programmer. The simulation of *a universe* from within could, for example, comprise a few galaxy clusters only or even just one galaxy hosting one intelligent species. According to current observations, *the universe* contains an estimated $1 \times 10^{12} - 2 \times 10^{12}$ galaxies [73]. If we, as internal programmers, were to simulate *a universe* because of constraints in computability, a reasonable approach would be to limit the number of astronomical objects. Also, if an external programmer's universe is infinite, a simulation of *a universe* could limit the complexity in terms of the objects contained and in size. Prevailing theories about *the universe's* evolution allow for the conclusion that its size is increasing and its rate of expansion is accelerating, but it is nevertheless finite. Finiteness supports the argument that *the universe* is really *a universe* as an external programmer's simulation is constrained by computational resources. Moreover, the recursion problem would result in less complexity per simulation down the hierarchy, as a simulation $s_x \in S = \{s_1, \ldots, s_n\}$ where $\lim_{n \to \infty}$ would be constrained by the computational resources provided by the computer in $s_{x-1}$ at hierarchy level $x - 1$ (see Fig. 3). The argument is an extension to the **first computability constraint**, finding application when simulating *a universe*, in which the computer is smaller than *the universe*, and potentially also smaller than *a universe* it simulates.

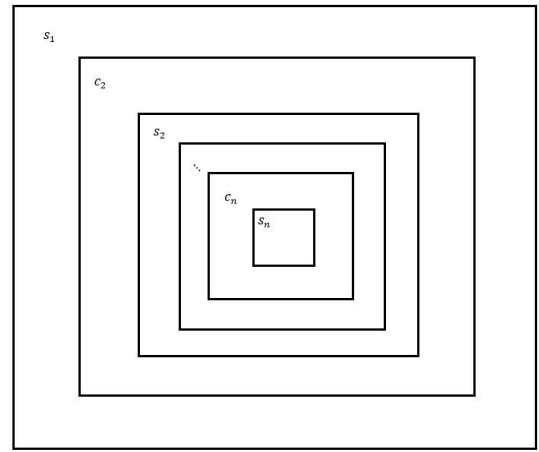

Fig. 3. A simulation hierarchy with $s_1$ marking *the universe*, which may be a simulation (or not). $c_{2\ldots n}$ are the computers running the simulations $s_{2\ldots n}$, each of which are smaller than the computers running them.

- The simulation of *a universe* from within or without is constrained by the same physical laws as a simulation of *the universe*. If an internal or external programmer decides to simulate parts of their universe, thus creating *a universe*, the physical predictability constraint prevents the evolution of the simulation of *a universe* to be identical to parts of *the universe*. The uncertainties in nature imposed by quantum physics allow for the simulation of *a universe* based on the same physical laws as *the universe*. The evolution of such a simulation could, in principle, be similar to *the universe* but need not be identical. For example, the mix of matter, anti-matter, and dark matter could be different, with less or more of each being present, or *a universe* in which anti-matter instead of matter makes up astronomical objects. If *the universe* contained more than one of each, its fate in the distant future, or even the past, would be different. Assuming that the simulation of

*a universe* results in a universal evolution similar to our observations, it is conceivable that an intelligent species emerges in this simulation. The second law of thermodynamics states that thermodynamic entropy - the state of disorder - in a closed or isolated system (such as *a universe*) will increase as this system propagates along the arrow of time. Contrary to the first assumptions, more thermodynamic entropy means more computational complexity and, consequently, more computational resources needed for simulating *a universe*. One way to think about thermodynamic entropy is the number of configurations a particle can assume: the more options a system has to arrange its particles, the more complex it is. Also, as a particle moves along the arrow of time, it interacts with other particles and entangles with other particles, which results in a more complex universe. Now, there is an intermediate step where interactions of components of *the universe*, for example, on an atomic or molecular level, decrease thermodynamic entropy locally and further increase computational complexity. Life is a good example; any living organism requires energy to function, reproduce cells, and maintain its structure. One way thermodynamic entropy affects living organisms is through cell degradation or cell death. The metabolism works against it by chemically transforming substances in living organisms' bodies, for example, transforming food into intermediate and end-products. These biochemical processes build up, break down and replace or maintain bodily substances and generate energy for energy-consuming activities, thus maintaining bodily functions and, therefore, life. Enzymes that catalyze chemical conversions are essential for metabolism. If foreign substances taken in from the outside are converted, one also speaks of foreign substance metabolism. The conversion of substances foreign to the organism into substances native to the organism is called assimilation. The opposite is dissimilation, the degradation of the organism's substances. Metabolism also includes the conversion of harmful substances into excretable substances. Metabolic processes can physically be interpreted as an exchange of free energy for order: Living organisms increase their order and consume energy in the process. In the organism, the thermodynamic entropy decreases, while in the environment, it increases. Therefore, even though the thermodynamic entropy of both *a* and *the universe* increases globally over time, the computational complexity keeps increasing due to an increasing number of interactions of the particles, and as long as complex structures such as astronomical objects and/or life emerge and are maintained locally. Assuming an intelligent species in the simulation is, at some point, capable of manipulating galaxies and/or larger structures in *the universe*, the increase in computational complexity will be even more significant. According to this argument, a computer used to simulate *a universe* would see an increase in computational demand as the simulation moves forward in time. The same is true for simulations in the simulation, also supported by the two recursion arguments introduced in this section. A few words on the arrow of time: we can consider three arrows of time [74] - the psychological arrow of time, the thermodynamic arrow of time, and the cosmological arrow of time.

– The psychological arrow of time is related to our own perception as to how the time in *the universe* and locally moves forward, or our memories.
– The thermodynamic arrow of time refers to the increase of thermodynamic entropy in *the universe*. Even though thermodynamic entropy may decrease locally, it always increases globally. No matter *the universe's* fate (see chapter V), thermodynamic entropy does not decrease. If the expansion continues forever, we expect to see the heat death of *the universe*, or the Big Chill [75]. If *the universe* collapses again, called the Big Crunch [76], thermodynamic entropy also does not decrease, as it keeps increasing even as space-time collapses.
– The cosmological arrow of time refers to the direction of *the universe's* expansion. It may be linked to the thermodynamic arrow of time if *the universe* continues to expand forever. This arrow of time would reverse would *the universe's* fate be the Big Crunch.

The presented argument will be referred to as the **third computability constraint**, as even in the simulation of *a universe* with no further simulations down the hierarchy, the computational complexity of the simulation would increase due to an increasing number of particle interactions along the arrow of time, and the energy-induced decreases of local entropies. Even if the external programmer's universe and the computer running the simulation are governed by different physical laws and the simulation is computationally simple when it starts, the computational complexity will increase and, given sufficient time, exhaust computational resources. Let it be said that external time outside the simulation does not need to run at the same pace as internal time within the simulation: even if the simulation runs at increased speed or is sped up from an outside perception to the inside time perception, this will not make a difference. Depending on *the universe's* fate, we may see different evolutions of thermodynamic entropy. If *the universe* continues to expand forever, thermodynamic entropy will continue to the point until no particle interactions can happen any longer, and thus, no additional degrees of freedom can be added to individual particles. Likewise, the global thermodynamic entropy will continue to grow in a big crunch scenario, but the shrinking surface bounding *the universe* may reduce complexity at the bounding surface, depending on its nature. If the computer used to run the simulation is bound to the same physical laws as the simulation, the **third computability constraint** can be delayed if the computer for simulating *a universe* is bigger than the size of *a universe*, which is, theoretically, possible, but not reasonable due to the complexity involved in building such a system.

- An extension to the physical and **computational predictability constraints** is the impossibility of separating the computer simulating *a universe* from within from *the universe*, even if the computer is not part of the simulation. The reason is quantum-vacuum fluctuations, which are the ability of space to create particles and their antiparticles seemingly out of nowhere. The foundation for the quantum-mechanical possibility of particle generation is Heisenberg's uncertainty principle (Eq. 4), which, in terms of energy, can be reformulated as Eq. 22, stating that the energy uncertainty multiplied by the time uncertainty must always be smaller than a natural constant in a small volume of space - the Planck's constant.

$$\Delta t \Delta E \geq \frac{\hbar}{2} \quad (22)$$

$$\frac{\Delta A}{\left|\frac{\partial \langle \hat{A} \rangle}{\partial t}\right|} \Delta E \geq \frac{\hbar}{2} \quad (23)$$

In Eq. 23, $\Delta t$ can be interpreted as the time that a quantum vacuum state changes significantly with respect to an observable $A$. This uncertainty principle is a direct consequence of the wave-particle duality and applies strictly to all elementary particles. Quantum fluctuations occur when a particle-antiparticle pair forms briefly without violating this uncertainty relation. For example, one consequence of quantum fluctuation is that there cannot be an absolute vacuum. Also, particle-antiparticle pairs are constantly created in "empty space" and disappear again. Experimental evidence for this situation is, for example, provided by the Casimir effect, where nearby metal plates exert forces on each other in a vacuum. Quantum fluctuations can also be used to explain the formation of particles in the vicinity of black holes, more precisely at their Schwarzschild radius. In this way, one of the two particles can escape from the black hole if the particle formation occurs in an area around this radius. In general, the internal complexity of black holes is not yet understood, as far as our current understanding goes, quantum information cannot be destroyed, so black holes should be tremendously complex internally. In certain cosmological models, the emergence of *the universe* in the Big Bang is also viewed as a quantum fluctuation. Now, if a computer is used to simulate *a universe* from within as part of *the universe* exactly, we would run in a mixture of the physical and **computational predictability constraints**: the computer's internal state would be influenced by the random emergence of particles caused by quantum fluctuations in *the universe*, preventing it from exactly predicting the future state of parts of *the universe*. From here on, this constraint is called the **combined predictability constraint**.

- An accurate simulation of *a universe* from both within and without may also result in the recursion problem outlined in the **first computability constraint**. If we, as external programmers, accurately simulate parts of *the universe*, say a number of galaxies or clusters, then whether we intend to simulate life is an important question. If life is part of the simulation - an assumption that seems reasonable given the simulation is accurate -, then the emergence of an intelligent species is probable. In case this happens, and the simulation is accurate, it stands to reason that the intelligent species in the subset-universe will also intend to simulate *the universe*. In the simulation of *a universe*, the simulation of the computer running the simulation is not required, and thus, the computer used for simulation does not cause recursivity as described in the **first computability constraint**, however, a species in the simulation intending to simulate *a universe*, will use a computer for carrying out the task, resulting again in a hierarchical set of simulations. The computer on the top level would have to provide the computational resources for computing a set of simulations $S = \{s_1,...,s_n\}$ where $\lim_{n \to \infty}$, resulting in the **first computability constraint** once a simulation at some level $n$ is run. All other constraints ignored, it can be argued that life in *the universe* may have been introduced by an external programmer instead of emerging by itself and that this would prevent recursion from happening. However, it stands to reason that an artificially introduced intelligent species would also contemplate simulating *a universe* and introducing an intelligent species in any such simulation, given that such a species does not emerge. Again, such a situation would result in the **first computability constraint**. The presented argument extends to the **first computability constraint** limited to simulating complex life.

The arguments presented in this chapter show that even if the intention is to simulate only parts of *the universe* exactly, several constraints prevent us from doing so. The arguments clearly show that no matter how powerful computers become, and no matter if the computers used for simulation are quantum computers, a simulation of *a universe* from within is not possible. However, there are no constraints preventing the simulation of *a universe* from without, thus, *a universe* that physically behaves like *the universe* but is not identical to it. Bringing all this together, experiments to verify whether we live in a simulation and whether an external programmer exists or not can be designed.

## VII. THE CREATOR-EXPERIMENTS

If an external programmer exists, and the external programmer created *the universe*, the external programmer can Be considered the creator, which, in different religions, assumes different forms. Religion won't be discussed any further in this context, as we are after verifying or refuting the simulation hypothesis. The primary goal of the outlined experiment is finding indications on whether *the universe* and everything in it, including humans and potentially other intelligent beings, were created by an external programmer and are part of a simulation. The hypothesis for the following experiments is plain and simple: "We do not live in a simulated universe and are not simulated beings." Before outlining the experiments, several assumptions made at the beginning of chapter VI are to be highlighted once more.

1) We have a complete understanding of the physical laws governing *the universe's* evolution and *the universe's* composition down to the physical level X. This assumption is certainly not true today.
2) The laws of quantum physics are the most fundamental physical laws. This assumption is footed on solid experimental ground, and we are fairly certain that this is true, at least for the strong and weak interaction and electromagnetism. Bringing together general relativity and quantum physics is a different story - as of the time this article was published, no theory of quantum gravity has been proposed that can be verified by experiment.
3) It does not matter if an external programmer decides to simulate *a universe* utilizing the same physical laws prevailing in their universe or a different set of physical laws. As soon as we act as external programmers and base our simulation on the physical laws prevailing in our universe, the actions of subsequent programmers in a potentially emerging simulation chain are candidate observations to look for in *the universe*.

In chapter VI, we outlined several constraints on creating a simulation of *the universe* and/or *a universe*, briefly summarized as follows:

- **Computational predictability constraint**: if the computer predicts the state of *the universe* at $t_1$ in $t_c$, it also predicts its own internal state at $t_r$. If the computer is asked at $t_c$ to predict the state of *the universe* at $t_2$, it will base its predictions on the state of *the universe,* including its own internal configuration at $t_1$. However, $t_c < t_1$, so the computer would start to change its internal configuration to predict *the universe*'s state at $t_2$ before $t_1$ has been reached and to its original predictions of its internal state at $t_1$ are not correct when $t_1$ is reached.
- **Physical predictability constraint**: it is impossible to simulate *the universe* entirely, as quantum physics imposes several constraints on the predictability of physical systems. Even if the computer carrying out the simulation would, in its internal state, use one particle of a certain type to simulate another particle of the same type, the uncertainty relation and the wave nature of particles prevent the computer from making exact predictions. Also, since there is no way to completely disentangle the computer's particles from the rest of *the universe's* particles in the simulation within, the computer changing internal state will always influence the state of the universe, thus making a simulation of *the universe* incomputable.
- **First computability constraint**: Every simulated computer would also have to simulate *the universe,* including itself, which is a recursion. The recursion is not only computationally expensive, it results in general incomputability of the simulation, as the available computational resources have to be used to compute a set of simulations $S = \{s_1, \ldots, s_n\}$ where $\lim_{n \to \infty}$ . Either the computer in $s_1$ would run out of resources at some point, or the computational resources in $s_n$ would not suffice to conduct another simulation $s_n + 1$. An extension of this constraint was introduced when it comes to simulating *a universe* and astronomical objects and/or life: If intelligent species tend to simulate *a universe* and all life in it, it is safe to assume that every simulated species will follow this tendency. Here, too, the occurring recursion would result in less complexity per simulation down the hierarchy, as a simulation $s_x \in S = \{s_1, \ldots, s_n\}$ where $\lim_{n \to \infty}$ would be constrained by the computational resources provided by the computer in $s_{x-1}$ at hierarchy level $x - 1$.[2]
- **Second computability constraint**: Running a simulation of *the universe* from within, a computer running would have to represent all particles of *the universe*, which we currently assume are $10^{78} - 10^{82}$, in its internal state. Building such a computer would require at least $10^{78} - 10^{82}$ particles; therefore, such a simulation is impossible.
- **Third computability constraint**: Even in the simulation of *a universe* with no further simulations down the hierarchy, the computational complexity of the simulation would increase due to an increasing number of particle interactions along the arrow of time and the energy induced decreases of local entropies, even as global thermodynamic entropy continues to increase. As the tendency of intelligent species in a simulation would be to create simulations including complex astronomical objects and life, the cumulative complexities of simulations down the simulation hierarchy would add up and consume all primary computing resources provided to the top-level simulation.
- **Combined predictability constraint**: Due to quantum vacuum fluctuations, the simulation computer's internal state would be influenced by the random emergence of particles caused by quantum fluctuations in *the universe* (of which the computer is part), preventing it from exactly predicting the future state of parts of *the universe*.

An experiment to confirm or refute that *the universe* and everything in it is not a simulation requires us to simulate *a universe* that physically behaves like *the universe*. Because of the computational, physical, and **combined predictability constraints** as well as the **first and second computability**

---

[2] There is no distinction made between life in a simulation and life as we know it, as we currently cannot know if we, ourselves, are simulated beings.

**constraints,** the simulation can't mimic even parts of *the universe* but only imitate it in terms of physical law, which, in our first and second assumptions, we claim to have understood completely and to the most granular level. Any accurate simulation of *a universe* must not start with an arbitrary state at an arbitrary time but with the beginning of space-time - the Big Bang, as of our current understanding.[3] One important aspect of such a simulation is that we can speed it up, concluding which emerging phenomena are statistically relevant and not coincidental appearances. The following observations may be made in such a simulation:

- Intelligent life emerges in the simulation: If in the simulation, intelligent life emerges, and it does so with statistical relevance over many runs of the simulation, we can conclude that life in *the universe* is no coincidence. However, the emergence of life alone is not conclusive as to whether *the universe* and everything in it is a simulation. If the intelligent life in the simulation does not intend to simulate *a universe*, we may use this as an argument in favor of not living in a simulation. If, however, the intelligent life in the simulation will simulate their universe, it is an indication of *the universe* and everything in it being a simulation, as the behavior is mimicked in each simulation down the simulation hierarchy $s_x \in S = \{s_1, \ldots, s_n\}$ where $\lim_{n \to \infty}$.

- Intelligent life does not emerge in the simulation: If in the simulation, intelligent life does not emerge, and it does not do so with statistical relevance over many runs of the simulation, we can either conclude that the emergence of life in *the universe* was a coincidence, or that life was artificially introduced by an external programmer. If life does not emerge, and if we appear as the external programmer and introduce intelligent life to the simulation, we may observe that the artificially introduced life intends to simulate their universe. If they do so, it is probable that in also their simulation, no intelligent life appears, and they artificially introduce it.

Both situations are inconclusive as to whether we live in a simulation, but we can use the observations to develop the experiment further. If an intelligent species in a simulation, no matter if they emerged or were artificially introduced to the simulation, creates a simulation of their universe, including intelligent life, this strongly indicates us being in a chain of simulations already. Now, each additional simulation draws computing resources from the level above it, and, applying the simplifying assumption that each simulation uses the maximum computing power available to it, each simulation further down the simulation hierarchy must follow the **first and third computability constraints** and feature reduced complexity and/or content. Moreover, due to the ever-increasing particle interactions along the arrow of time and local decreases of entropies in each simulation, each simulation comes with an ever-increasing demand for computational resources. Computing resources on every level, even the top level, must be constrained, either conditionally or by resources, except if the size of the computer is larger than the size of *a universe* it simulates. Even though each simulation down the simulation hierarchy is less complex than the previous one, each simulation $s_x \in S = \{s_1, \ldots, s_n\}$ down the simulation hierarchy increases in complexity over time as its global thermodynamic entropy increases and local entropies decrease, no matter the fate of the respective simulated universe. If we participate in a chain of simulations, the very distant future in our simulated time (which does not need to run at equal speeds to the external programmer's time) will halt the exhaustion of computational resources on any level and thus halt the increase in thermodynamic entropy on any level. If thermodynamic entropy (and thus time) is halted at any level, all entropies in simulations further down the hierarchy also immediately come to a halt. According to the **second and the third computability constraints**, the complexity in *the/a universe* will increase, no matter the fate of *the/a universe,* and exhaustion of computational resources is unavoidable if the computer is smaller than *a universe* it simulates, except the global thermodynamic entropy is reversed. As of our current understanding, it is not possible to globally reverse thermodynamic entropy in *the universe*, but if it were a simulation, an external programmer would certainly have the ability to do so, and we have the ability to do so if we were to run the simulation of *a universe* in our universe. There may be other ways of reducing complexity, for example, drastically reducing the number of objects in the simulation as global thermodynamic entropy increases. Now, sticking with the example of thermodynamic entropy reduction by an external programmer, if we were to simulate *a universe*, $s_1$, and in that simulation, intelligent life simulating *a universe*, $s_2$, emerges, the same computational constraints appearing in $s_1$ will appear in $s_2$. In this experiment, we, as external programmers, work against exhaustion of computational resources in $s_2$ - *the universe* we simulate - by reversing the global thermodynamic entropy of *a universe* we simulate, resulting in less complexity, an increasing temperature, less complex molecules, and more. A reversal of global thermodynamic entropy does not require local entropies to be reversed. In other words, decreasing global thermodynamic entropy does not require increasing local thermodynamic entropy. If we now observe that this also happens in $s_2$ for $s_3$, where $s_3$ is the simulation of *a universe* running in $s_2$, and further down the hierarchical chain of simulations, we may conclude that reversing the thermodynamic entropy in a chain of simulations to prevent an exhaustion of computational resources and keeping it alive is expected behavior. As the thermodynamic arrow of time prevents us from decreasing the global thermodynamic entropy in *the universe*, we do not have accepted theories of *the universe's* fate in such a scenario. If, however, we were to observe a decrease in global thermodynamic entropy inexplicable by physical law, this would allow for the conclusion that we live in a simulation controlled by an external programmer.

---

[3] If, at some point in the future, new insights lead to another cosmological model, the Big Bang model may very well be replaced by that one.

Thermodynamic entropy reduction to reduce computational complexity is merely one conceivable example of actions conducted by an external programmer that may lead to observable consequences in a simulation. It is conceivable that an external programmer conducts even more severe changes in fundamental physical laws, such as compactifying spatial or temporal dimensions. In physics, one of the most formidable problems is the search for a theory of everything that explains all physical phenomena from the subatomic to the cosmological scale. One candidate for a theory of everything is M-theory [77–79], consolidating the string theories [80]. More specifically, M-Theory is an attempt to extend and generalize string theory and is the eleven-dimensional unified theory of the five string theories and supergravity. It was shown that the five string theories are just special limiting cases of M-theory [81]. Even today, as M-theory is a subject of active research and cannot be mathematically described, it already reveals remarkable properties of strings, space-time, and *the universe*. Besides unifying the four fundamental forces, it reconciles quantum physics with general relativity. One of the most far-reaching consequences arising from it is the realization that the fundamental building blocks of *the universe* are not exclusively one-dimensional strings but multidimensional objects, so-called branes. Our perception and current experiments lead us to describe the macroscopic universe utilizing four-dimensional space-time, which is where we see a disconnect with M-theory, which features 10 spatial dimensions and one time dimension. The seven unobserved spatial dimensions are thought to be curled up on the Planck scale, which current particle accelerators cannot access. Today's particle accelerators produce energies around $10^{13}$ eV, but the Planck energy is 1.2 $\times 10^{19}$ GeV. It was shown that supergravity not only permits up to eleven dimensions but is most elegant in this maximal number of dimensions, which is how the extension of the common 10-dimensional space-time of the string theories to the 11-dimensional space-time of M-theory came about. Now, assuming that at the Big Bang *the universe* featured 11 macroscopic space-time dimensions, an external programmer could use compactification (Fig. 4) to decrease computational complexity as thermodynamic entropy increases along the arrow of time.

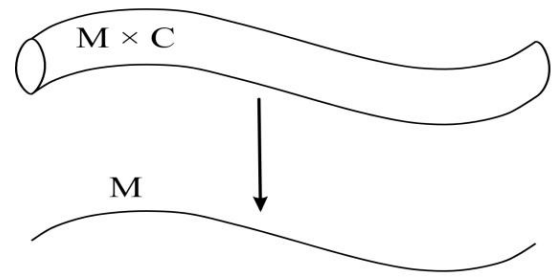

Fig. 4. Compactification of the space $M \times C$ over compact $C$

Some of the extra dimensions are assumed to form circles, or close up on themselves, which would reduce theoretically infinite or very large dimensions to dimensions of finite length. One common assumption as to why some space-time dimensions may be compactified is that cosmic inflation scaled up only some of the dimensions and curled up others in circles on the Planck scale, with reasons unknown. If true, the four forces we perceive in our 4-dimensional space-time are merely manifestations of the same unified force of a higher-dimensional space-time. Naturally, as physical systems tend to simplicity, one unified force would be more "reasonable" than four distinct forces. That being said, it is reasonable to assume that an external programmer would have the ability to compactify or even eliminate space-time dimensions, whereby the complete elimination of dimensions would, presumably, be more catastrophic. What's more, the compactification to 4-dimensional space-time may be local, whether it is an inherent feature of the universe or caused by an external programmer. Other space-time bubbles may very well be higher- or lower-dimensional. Should humans ever detect higher-dimensional space-time regions, and should these be primarily located in areas not populated by complex astronomical objects, this may indicate that space-time has been artificially compactified in other regions and we participate in a simulation chain.

Another form of dimensionality reduction for delaying the **second computability constraint** is the holographic universe [58, 59]. It is the hypothesis that for every description of the dynamics of a space-time area, there is an equivalent description that is only localized on the edge of this area. As a result, the maximum possible entropy of a region of space does not depend on its volume but on its surface only, as in the case of the Bekenstein-Hawking entropy of black holes [54, 57]. The holographic principle provided an interpretation of black hole entropy and was even motivated by it. Considering gravity, the information content - the number of possible arrangements of particles and fields - cannot be a purely local quantity because it would be proportional to the volume.[4] The surface area of a black hole's event horizon - the boundary surface of the black hole formed by the Schwarzschild radius is a direct measure of the entropy or the information content of the enclosed volume of space and, thus, the masses contained therein. A black hole always represents the maximum possible concentration of matter in a region of space and, hence, the upper limit of possible entropy or information in the volume of space it occupies. The holographic principle postulates that any information that exceeds the surface area of a black hole's event horizon is completely encoded on the interface spanned by the Schwarzschild radius, similar to a two-dimensional hologram that contains three-dimensional image information. Because the Schwarzschild radius of a black hole is directly proportional to its mass, the encodable volume grows faster than the surface area. To encode four times the volume, only twice the surface is available. That means that the information density of a region of space decreases with increasing volume, just as the average mass density of a black hole decreases analogously with the size, or, more concisely, that information equals surface. That, in turn, allows for the conclusion that, for example, our 4-dimensional spacetime, including its physical laws, could be encoded on its 3-dimensional boundary. It was shown that a theoretical universe described by superstring theory in an anti-de-Sitter spacetime [82] is equivalent to a quantum field theory operating on its boundary [83]. First, this was confirmed for the 5-dimensional anti-de-Sitter space-time [84], and it indicates that it is impossible for beings in *a universe* to determine if they exist in a 4-dimensional universe operating on a quantum field theory or a 5-dimensional universe operating on a string theory. An external programmer could potentially use this fact and encode a more complex, higher-dimensional universe as physical theories operating on its lower-dimensional boundary surface or vice-versa (which is not a hologram). The latter case could be particularly interesting for reducing computational complexity, as only the physical laws could be encoded into a higher-dimensional spacetime. The *universe* and its evolution could happen on its lower-dimensional boundary surface. An external programmer could define physical laws, and rules for evolution could be defined in the higher-dimensional space-time and projected onto the boundary surface, where they govern the evolution of a lower-dimensional universe. If experiments [85] show that the holographic universe is true, it may be interpreted as an indication of us participating in a simulation chain.

---

[4] The term holographic is based on the analogy to the hologram, which stores a three-dimensional image on a two-dimensional photo plate.

Certainly, there are many more drastic ways an external programmer can think of to reduce the universe's computational complexity, and not all of them may be detectable. Summing up, we may observe the following scenarios when conducting simulations of *a universe* (all of them executed many times to ensure the statistical significance of the results):

- Intelligent life emerges in our simulation of *a universe* and further down the simulation hierarchy: If that is the case, we may reason we live in a chain of simulations, as the simulation of *a universe* including intelligent species intending to simulate *a universe* seems normal behavior. In this scenario, further observations may be made:
  - The observation of fundamental and grave physical interventions by external programmers in a chain of simulations, such as a reversal of the global thermodynamic entropies, compactification of dimensions, the implementation of the holographic principle in various forms, an intentionally caused big crunch, or similar. Observing such dramatic interventions, we may conclude that we participate in a simulation chain.
  - The observation of a reversal of the thermodynamic entropies, or compactification of dimensions, the implementation of the holographic principle in various forms, an intentionally caused big crunch, or similar in a chain of simulations executed many times is not statistically relevant: we are unique in implementing such drastic physical interventions in our simulation $s_1$, which allows for the conclusion we do not participate in a simulation chain or we are the first ones in a simulation caring about exhaustion of computational resources of simulations down the simulation hierarchy. If all intelligent life down the simulation hierarchy behaves differently than we do in the sense of thermodynamic entropy reversal (or the like), we can conclude we are the first external programmers and do not live in a simulation ourselves.[5]
- No further simulation hierarchy emerges despite intelligent species emerging in our simulation of *a universe*: From this scenario, we may conclude that we are the first external programmers and do not live in a simulation ourselves.
- No intelligent life emerges in our simulation of *a universe* despite all other physical behavior in the simulation being identical to *the universe*: From this scenario, we may either reason that we are the first ones conducting a simulation and the emergence of life as we know it was a coincidence, or that an external programmer introduced life into *the universe* and we live in a simulation. Now, if we conduct a simulation of *a universe* ourselves and observe the emergence of a simulation hierarchy once we artificially introduced intelligent life and none of the programmers down the hierarchy have to introduce it because it emerges by itself, we may conclude we do not live in a simulation and are the first ones simulating *a universe*. If a simulation hierarchy emerges with simulated intelligent life introducing intelligent life themselves artificially in each new simulation, we may conclude we are in a chain of simulations.

---

[5] We do not include other "different" behavior in our considerations, as emerging life may be intelligent, but otherwise completely different in structure and behavior than humans.

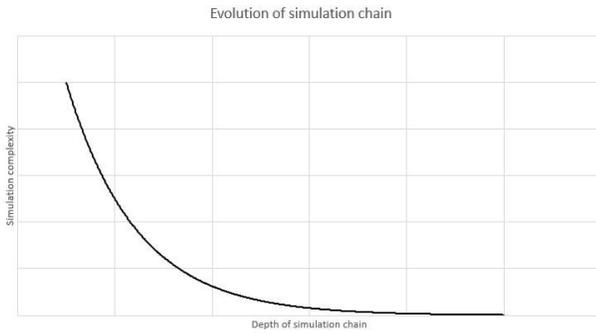

Fig. 5. Further down the simulation chain, the complexity of each new simulation is less than the previous simulation, as the simulation $s_{x-1}$ will, next to running itself, include the computer used for running $s_x$.

If we observe a scenario in which intelligent species and a simulation hierarchy emerges (with or without our doing), based on the **first and third computability constraints** we know that the complexity of every simulation down the hierarchy will increase over time. Also, we know that each simulation down the hierarchy will be less complex than the previous one - $s_{x+1}$ is less complex than $s_x$, as in $s_x$, a computer simulating $s_{x+1}$ with equal complexity than $s_x$ would need to consist of at least as many particles as $s_x$ itself (Fig. 5). At some point down the hierarchy, the last simulation $s_n$ will emerge, as the complexity involved in creating further simulations cannot be achieved due to insufficient complexity of $s_n$. We can also start with a simple simulation and increase complexity. By doing so, we determine which is the first simulation of sufficient complexity for intelligent life simulating *a universe* to emerge. By comparing the two simulations, the second to the last one in the simulation chain - $s_{n-1}$, in which an intelligent species is still simulating *a universe* $s_n$ -, and the first one of sufficient complexity to host an intelligent species simulating *a universe* $s_{min}$, we may observe and conclude the following:

- The two universes share many similarities in their evolution, structure, and content: We observe that it does not matter if we let the simulation chain evolve down to the least complexity or if we start a simulation with the least complexity needed for intelligent life to emerge simulating *a universe* directly - the least complex universes are identical. Assuming that $s_{min}$ is similar to *the universe,* and both are similar to each other, $s_{n-1}$ is also similar to *the universe*. From that, we may conclude that the behavior of our simulation chain is correct. On this result alone, we may not be able to conclude whether we participate in a simulation chain.
- The two universes are very different in evolution, structure, and content: Again, assuming that $s_{min}$ is similar to *the universe* but both are dissimilar to each other, $s_{n-1}$ is dissimilar to *the universe*. Dissimilarity does not exclude the emergence of intelligent life simulating *a universe*. The result is inconclusive, as we may consider one or more of the following:
    - Down the simulation chain, an evolution happened to cause the dissimilarity.
    - Our simulation chain does not correctly simulate *a universe*, and the error propagates down the simulation chain.
    - Each simulated universe and *the universe* (simulated or not) are special, as down the simulation chain, different behavior emerges in each of the simulations even though the simulations are physically correct.
    - On the dissimilarity alone, we may not be able to conclude whether we participate in a simulation chain.

## VIII. Conclusion

Proving that we do not live in a simulation is a complicated endeavor grounded in computer science, physics, and philosophy theory and practice. But, as always, crucial scientific evidence isn't to be found purely on theoretical grounds but also through experiments and observations. The question of whether we live in a simulation is open. With the experiments, constraints, and proposed observations outlined, we hope to obtain indications in favor of or against the simulation hypothesis based on our foundation of perceived and measurable reality.

The six outlined constraints, built on theoretical computer science and physics, experimental evidence, logic, and observations, define the boundaries within which a universal simulation can be conducted, such as the impossibility of simulating *the universe* at full scale from within. The constraints also prevent an external or internal programmer from building a computer capable of precisely simulating parts of *the universe* and thus from constructing a simulation identical to it, which - if feasible - would allow for the prediction of the future. Whereas Alan Turing showed what is computable, we outline which computers are constructible and how an external or internal programmer may use the ultimate computational resources to simulate *a universe*.

We are all thinkers of our time, and philosophy today is often a matter of taking Richard Feynman's famous "chalk quote" from one of his classes - "What I cannot create, I do not understand" - literally: It is a thinking practice. Many humans agree on fundamental beliefs and/or assumptions, such as mathematics being *the universe*'s syntax and semantics, in which the physics underlying and governing it can be expressed. More abstract ideas grounded in solid scientific theory and/or evidence state that while quantum physics is the "machine language" of *the universe*, many interpretations are possible and rigorous, such as particles' constantly collapsing wave functions resulting in an infinite number of parallel universes [9]. Quantum physics aside, it is conceivable that we exist in a space-time bubble 13.8 billion light-years in diameter governed by different physical laws than other spacetime bubbles in *the universe* [86]. Every physical theory rests upon the shoulders of giants, encompassing a myriad of ideas produced over time and countless questions that require further probing. Something inexplicable - mysterious even, in some instances - often provides the grounds for human endeavor and the search for answers. In this paper, we have argued from an "It" perspective there is something to be discovered about reality, independent of the physical laws governing it. We have proposed experiments and observations regarding our perceived and measured physical reality and the computability within such a formal system. Physics has always been about pushing beyond what is known and understood. One trend taking many forms is abandoning established physical laws and theories in favor of finding the new physics of *the universe*. In this paper, however, we have taken a clear stance in favor of well-established physical laws, models, and theories that, while not experimentally proven, seem "reasonable" based on our understanding of how *the universe* functions.

The authors of this paper recognize that theories stating consciousness is fundamental and that reality is an illusion building on ancient philosophy in which space-time is more like a headset [87]. Such theories state that our experiences are real but are grounded in the notion that consciousness is the foundation of our perceived reality and that physicalism, thoughts, and objects can emerge. If consciousness is fundamental and space-time and physical laws and mathematical structures can emerge organically - and if the provability of the laws of quantum mechanics can be projected onto such a reality - then halting and/or reversing entropy, for example, must be within the scope of an external programmer's power. However, even considering the most fundamental assumption of such a theory - universal consciousness -, there is no attempt to argue for or against it in this paper, as this was left open for follow-up studies.

The proposed experiments are based on several assumptions, which compose our tentative understanding of natural laws today, including the laws of quantum physics, which correspond with the most fundamental physical laws. In this paper, we have argued that certain limitations exist within the current scope of computational physics and our understanding of how a simulation hypothesis might be brought to life within our current understanding of quantum theory. Yet, we still argue along the lines of Gödel's incompleteness theorem and the Halting problem outlined by Alan Turing that progress is always possible. When it comes to the philosophical question of creation - be it from simplicity to complexity or from digital entities (the bit) to the physical perceived reality (the "It") - there is always an underlying presumption of "something" existing. Within any formal system, there has to be a starting point or a limitation from which progress follows. Without assuming at least one precondition, a scientific venture is impossible. Therefore, this paper aims to spark a discussion about incorporating the outlined constraints and limitations into solid theories and experiments instead of arguing for a final solution to the simulation hypothesis. This experience also proves that, although reductionism as a fundamentally scientific approach might not lead to any conclusive theory underlying reality, it operates within the given framework. There is still a tremendous amount of work regarding many of today's accepted narratives. In various forms, quantum physical laws lead to the **physical predictability constraint**, through which the most fundamental laws prevent an exact simulation and, thus, a prediction. Proponents of a deterministic universe strongly oppose the non-determinism of quantum physics some of the outlined constraints are built upon, with two of the most prominent proposals for determinism being superdeterminism and deterministic quantum mechanics. We logically reason against both theories and show why these are insufficient arguments for a deterministic universe and against free will. Pondering simulations is nevertheless fruitful, as other avenues than simulating *the universe* exactly are wide open to exploration. The simulation of *a universe* - a simulation smaller than the computer used to run it and governed by the same physical laws as *the universe* is possible. In such a simulation, intelligent life - and here we point out that our definitions of both intelligence and life are certainly incomplete and that life, in general, doesn't necessarily need to be similar to human life in composition and behavior to qualify as such - simulating *a universe* may emerge or be artificially introduced by an external programmer, which is the foundation for the simulation chain experiment described herein. Suppose a simulation chain emerges in such a simulation. In that case, this doesn't necessarily invite the conclusion that we are already participating in a simulation chain ourselves, meaning that *the universe* is a simulation. As outlined in the **third computability constraint**, each simulation down the simulation chain will be less complex than the previous one, as the latter must include the computer running the former. What's more, each of the simulations, no matter the complexity compared to other simulations in the chain, will become computationally more complex over time, as global thermodynamic entropy always increases, even as local thermodynamic entropies may decrease when, for example, astronomical objects form, or life emerges. Increasing thermodynamic entropy increases complexity as particle interactions, and the number of configurations particles can assume increases along the arrow of time. Increasing complexity inevitably increases computational complexity and will draw more computational resources. As all computational resources are finite, any time a simulation computer's resources are exhausted, all simulations down the chain come to a halt. We propose ideas as to how an external programmer can temporarily circumvent the exhaustion of computational resources, some of which may be detectable in *the universe* by way of experiment or observation. Nothing prevents an external programmer from fundamental and grave physical interventions in a chain of simulations, such as a reversal of global thermodynamic entropy, the compactification of dimensions, the implementation of the holographic principle in various forms, or an intentional big crunch. Suppose we observe one or more of these scenarios - or others with the same severity - occurring many times down the simulation chain and, thereupon, conclude that the related events are statistically relevant. In that case, we may derive that this is expected behavior, and should we find indications for it in *the universe* as well, we may further conclude that we participate in a simulation chain. Other scenarios outlined consider no emergence of a simulation chain or intelligent life intending to simulate *a universe* once we simulate *a universe* ourselves, which allows for different conclusions as to whether we exist in a simulation. Lastly, one of the experiments outlined concerns starting a simulation of sufficient complexity such that intelligent life intending to simulate *a universe* emerges and comparing this simulation with the least complex one in a chain of simulations in which intelligent life still intends to simulate *a universe*. If these are similar, we can conclude that the simulation chain we created is correct, but it also means that our current knowledge base is inconclusive and demands further investigation. Sixty years after Asimov's "last question", we are confronted with the doom declaration of space and time, various views of possible multiverses, scattered interpretations of quantum mechanics, and a world in which mathematics is taken to the spiritual cathedral of many a belief system. The very notion of creation remains an open philosophical question yet still paves the way for scientific progress through further experience and advancement in quantum technologies.

In other words, there is still room for one final question. We look forward to your reflections and an exciting discussion: let there be light.